\documentclass[twocolumn]{aastex61}
\pdfoutput=1 
\usepackage{amsmath,amstext}
\usepackage[T1]{fontenc}
\usepackage{apjfonts} 
\usepackage[figure,figure*]{hypcap}
\usepackage{xspace}
\usepackage{gensymb}
\usepackage[normalem]{ulem}

\newcommand{\Ha}{H$\alpha$\xspace}
\newcommand{\pp}{P11695\xspace}

\defcitealias{Poggianti2017a}{Paper I}
\defcitealias{Fritz2017}{Paper III}
\defcitealias{Gullieuszik2017}{Paper IV}

\shorttitle{Gas accretion onto an isolated galaxy}
\shortauthors{B. Vulcani et al.}

\begin{document}

\title{GASP VII. Signs of gas inflow onto a lopsided galaxy}

\author{Benedetta Vulcani}
\affiliation{School of Physics, University of Melbourne, VIC 3010, Australia}
\affiliation{INAF- Osservatorio astronomico di Padova, Vicolo Osservatorio 5, IT-35122 Padova, Italy}
\author{Bianca M. Poggianti}
\affiliation{INAF- Osservatorio astronomico di Padova, Vicolo Osservatorio 5, IT-35122 Padova, Italy}
\author{Alessia Moretti}
\affiliation{INAF- Osservatorio astronomico di Padova, Vicolo Osservatorio 5, IT-35122 Padova, Italy}
\author{Michela Mapelli}
\affiliation{INAF- Osservatorio astronomico di Padova, Vicolo Osservatorio 5, IT-35122 Padova, Italy}
\author{Giovanni Fasano}
\affiliation{INAF- Osservatorio astronomico di Padova, Vicolo Osservatorio 5, IT-35122 Padova, Italy}
\author{Jacopo Fritz}
\affiliation{Instituto de Radioastronom\'ia y Astrof\'isica,
UNAM, Campus Morelia, A.P. 3-72, C.P. 58089, Mexico}
\author{Yara Jaff\'e}
\affiliation{European Southern Observatory, Alonso de Cordova 3107, Vitacura, Casilla 19001, Santiago de Chile, Chile}
\author{Daniela Bettoni}
\affiliation{INAF- Osservatorio astronomico di Padova, Vicolo Osservatorio 5, IT-35122 Padova, Italy}
\author{Marco  Gullieuszik}
\affiliation{INAF- Osservatorio astronomico di Padova, Vicolo Osservatorio 5, IT-35122 Padova, Italy}
\author{Callum Bellhouse}
\affiliation{University of Birmingham School of Physics and Astronomy, Edgbaston, Birmingham, England}
\affiliation{European Southern Observatory, Alonso de Cordova 3107, Vitacura, Casilla 19001, Santiago de Chile, Chile}

\begin{abstract}
Theoretically, inflowing filaments of gas are one of the 
main causes of growth for a galaxy. Nonetheless, observationally, probing 
ongoing gas accretion is challenging. As part of the Gas Stripping Phenomena in galaxies with MUSE (GASP) program, we present the analysis of a spiral galaxy at $z=0.04648$ whose characteristics indeed are consistent with a scenario in which gas accretion plays a major role. The most salient indirect parts of evidence that support this picture 
are: 1) The galaxy is  isolated, its position rules out the mechanisms expected in dense environments. 2) It shows a pronounced lopsidedness extending toward  West. According to the spatially resolved star formation history, this component was formed  $<6\times 10^8$ yr ago. 3) It has many large and elongated HII regions that are indication of a fragmentation due to disk instability. 4) The stellar and gas kinematics are quite symmetric around the same axis, but in the gas the locus of negative velocities shows a convexity toward East, as if new gas has been infalling with different orientation and velocity. 5) The metallicity distribution is  inhomogeneous and shows exceptionally steep gradients from the center toward the outskirts, especially in the South-West side. 
6) The luminosity weighted age is generally low ($\sim 8$ Gyr) and particularly low (<7 Gyr) along a trail crossing  the galaxy from South-West toward North. It might trace the path of the accreted gas.
These findings point to  an inflow of gas probably proceeding from the South-West side of the galaxy.
\end{abstract}

\keywords{galaxies:general --- galaxies:intergalactic medium --- galaxies:evolution --- galaxies:kinematics and dynamics}

\section{Introduction}
A large fraction of galaxies in the local Universe present a non-symmetric light distribution in their disk, whose spatial extent is much larger along one half of the galaxy than the other \citep{Baldwin1980, Block1994, Richter1994, Rix1995, Schoenmakers1997, Zaritsky1997, Haynes1998, Matthews1998, Swaters1999, Bournaud2005}. The asymmetry is  often visible also in the rotation curves, such that the shape and the maximum velocity are different in the two halves of a galaxy \citep[e.g.][]{Mihalas1981, Jog2002}. The percentage of these galaxies depends on the environment: it is larger than 50\% in  Hickson compact groups \citep{Rubin1991, Nishiura2000}, and is $\sim25\%$ in the field \citep{Rubin1999, Zaritsky1997, Sofue2001}. 
Given the frequency of this phenomenon, the lopsided modes should be fairly long-lived \citep[e.g.,][]{Baldwin1980} or excited frequently. 

Generally, the lopsidedness is clearly visible in the gaseous component (HI), and occurs also in the central regions, where the nucleus is offset with respect to the outer isophotes.  
The asymmetry is of type $m=1$, where $m$ is the azimuthal wavenumber \citep[e.g.][]{Bournaud2005}.

A similar fraction of galaxies ($\sim 50\%$) also present deviations from the single plane and exhibit an integral-shape with typical amplitudes of a few degrees in the stellar disc \citep[e.g.][]{Sanchez1990,  Reshetnikov1998, Ann2006}. These galaxies usually also present even stronger warps in the neutral hydrogen \citep[e.g.][]{Sancisi1976, Bosma1981, Briggs1990, Bottema1995}.

The asymmetry might be the result of different processes, such as tidal interactions \citep{Beale1969}, mergers \citep{Walker1996}, satellite gas accretion \citep{Zaritsky1997}, lopsided dark matter halo \citep{Jog1997, Jog2002, Angiras2007}, off-center disk in halo \citep{Levine1998, Noordermeer2001},  gas accretion \citep{Bournaud2005} or, in clusters and groups, ram pressure stripping \citep[e.g.,][]{Young2006}.  In addition to these externally triggered processes, the disk lopsidedness could also arise as a global instability in self-gravitating disks \citep{Junqueira1996, Bournaud2005}. As discussed by \cite{Mapelli2008}, the different scenarios might not be exclusive and different processes could  induce different degrees of lopsidedness in the different galaxy components.  For example, ram pressure might create only moderate tidal gas tails in galaxies without inducing lopsidedness in the stellar population; galaxy interactions in the form of flybys might account for much stronger asymmetries simultaneously both in the gaseous and in the stellar disc;  gas accretion from filaments might produce even more pronounced lopsidedness, first in the gaseous disc, and only at later times in the stellar component, due to star formation in the asymmetric gaseous disc.

Several theoretical mechanisms have been proposed to explain the formation and maintenance of warped discs \citep[e.g.][and references therein]{Binney1992, Kuijken2001, Sellwood2013}. Among the proposed scenarios, discrete modes of bending in a self-gravitating disc \citep{Toomre1983, Sparke1988}, misaligned dark halos \citep{Dubinski1995}, galaxy interactions and accretion of satellites \citep[e.g.][]{Huang1997, Schwarzkopf2001, Kim2014}, direct accretion of intergalactic matter in the outskirts of galaxies \citep[e.g.][]{Revaz2001, vanderKruit2007, Roskar2010}, extragalactic magnetic fields \citep{Battaner1990},  cosmic infall and outer gas accretion \citep[e.g.]{Binney1992}, and others. This large variety of proposed mechanisms and their modifications probably indicates that there is no single mechanism responsible for all observable warps in galaxies. The current situation looks like the largest warps are mostly caused by tidal distortions \citep{Schwarzkopf2001, Ann2006}, whereas relatively small warps are triggered and supported by a variety of mechanisms.

Overall, the most promising processes to describe lopsidedness are  tidal encounters and gas accretion, with the former being the dominant mechanism for group galaxies. Indeed, a perturbation due to a tidal encounter between two galaxies with an arbitrary orientation can generate a force term, which can then induce lopsidedness in the galaxy \citep{Combes2004}. Lopsidedness can also be generated more indirectly due to the response of the disk to the distorted halo which feels a stronger effect of the interaction \citep{Weinberg1995, Jog1997, Schoenmakers1997}. A generally stronger perturbation resulting from the infall of a satellite galaxy can also result in the disk lopsidedness as shown in the N-body simulation study by \cite{Walker1996}, who suggest that minor mergers can induce lopsidedness over a long time-scale ($\sim$1 Gyr).

While tidal encounters can explain the observed amplitudes of disk lopsidedness, the N-body simulations by \cite{Bournaud2005} show that they cannot explain  various observed statistical properties, such as the higher lopsidedness seen for the late-type field galaxies. 
Indeed, tidal encounters and mergers would tend to lead to the secular evolution of a galaxy towards earlier-type morphologies. Thus, if tidal interactions were the primary mechanism for generating lopsidedness, then the early-type galaxies should show a higher amplitude of lopsidedness. In addition, the amplitude of the lopsidedness seems not to correlate with the strength of a tidal encounter \citep{Bournaud2005}, or with the presence of nearby neighbors \citep{Wilcots2004}.  Tidal encounters  typically generate a fast mode which is expected to be not long-lived 
\citep{Ideta2002}.

Thus  mechanisms such as gas accretion from outside the galaxy need to be invoked to explain  the lopsidedness in field galaxies.

There is growing evidence that galaxies steadily accrete gas from the external regions, as seen from cosmological models \citep{Semelin2005}, and also observed in nearby galaxies \citep{Sancisi2008}.
 
Smoothed particle hydrodynamics (SPH) simulations by \cite{Keres2005} and grid-based adaptive mesh refinement simulations \citep[e.g.][]{Ocvirk2008}  predict that galaxies can accrete large amounts of gas from the Inter Galactic Medium (IGM) along cosmological filaments and
show that a fraction of the inflowing gas  is delivered to the disk not heated to the virial temperature of the halo  \citep[the so-called ``cold accretion mode'' - e.g.,][]{Keres2005, Dekel2006, Ocvirk2008, Brooks2009,  Dekel2009,  Faucher-Giguere2011, Faucher-Giguere2011b,  vandeVoort2011, Hobbs2015,Silk2012, Genel2012}. The cold accretion mode appears to be most efficient at high redshifts ($z > 2$), but at low redshifts it seems  to mainly occur in low-mass galaxies with M$_{halo}$ $\leq 5 \times 10^{11} M_\odot$. 

However,  more recently,  \cite{Nelson2013}  showed that the role of cold accretion might have been overestimated in previous smoothed particle hydrodynamics simulations. The new calculations, performed with the moving mesh code \citep[{\sc arepo,}][]{Springel2010}, indicate that the majority of gas is heated to the virial temperature before accreting onto the halo, at all galaxy masses \citep[see also][]{Torrey2012, Nelson2016}.

The cold accreted gas does not necessarily mix with the existing gaseous halo and over a few billion years it can represent a significant fraction of the disk mass itself (both as gas phase and as newborn stars).  This accretion can be asymmetrical \citep[e.g.][and references therein]{Jog2009}, due to the different gas distribution along the filaments and to the different accretion rate.

The degree of lopsidedness decreases after a few Gyrs, even if accretion is not stopped, because the disk re-orients itself in regard to the filament. If there are more filaments with different accretion rates  \citep[e.g.,][]{Semelin2005}, 
the lopsidedness will generally be more moderate.
In the simulations,  when accretion is stopped  the lopsidedness has a typical lifetime of 3 Gyr \citep{Bournaud2005}. 

Therefore gas accretion can be the clue to explaining strong lopsidedness in galaxies that have no sign of recent interaction/merger.

Gas accretion might also induce the formation of clumps in the disk due to local gravitational instabilities. Clumping instabilities also cause starbursts, with a star formation rate typically enhanced by a factor of 10 in the present models. These bursts are thus not related to interactions, although the large-scale environment should influence this kind of activity indirectly through its role in the growth of disks \citep[e.g.][]{Noeske2007, Elmegreen2007}.
 
Finally, another consequence of gas accretion might be metallicity gradients. This connection is poorly explored in the literature. Primordial gas accreted through filaments should be characterized by a lower metallicity than that of the target galaxy. Note that instead both flybys and ram pressure are expected not to significantly affect metallicity gradients \citep{Mapelli2008}. 
However, it is important to keep in mind  that stars form from a mixture between the gas in the cold filament and the gas which was already in the disk.
Moreover, if the gas accretion from the filament is not recent, but occurred in the past, phenomena like pollution from supernovae, stellar winds and other feedback mechanisms might have already erased any metallicity gradient both in the gaseous and in the stellar component.

In the local Universe, the advent of wide-field integral field spectroscopy (IFS) is permitting to study metallicity gradients in a statistical way \citep[e.g.][]{Sanchez2014, Ho2015}. On-going large IFS surveys include the Calar Alto Legacy Integral Field Area (CALIFA) Survey \citep{Sanchez2012}, the Sydney-AAO Multi-object Integral field spectrograph (SAMI) Survey \citep{Croom2012}, the Mapping Nearby Galaxies at Apache Point Observatory  (MaNGA) Survey \citep{Bundy2015}. Overall, disc galaxies have been found to universally exhibit negative metallicity gradients, i.e. the centre of a galaxy has a higher metallicity than the outskirts \citep[e.g.][and references therein]{Zaritsky1994, Moustakas2010, Rupke2010b, Sanchez2014}.

Gas accretion and metallicity gradients have been also studied at higher redshift. For example, \cite{Cresci2010} observed low metallicity regions in three galaxies at $z\sim$ 3,  providing  evidence for the actual presence of accretion of metal-poor gas in massive high-z galaxies, capable to sustain high star formation rates without frequent mergers of already evolved and enriched objects. 
In this paper we present the analysis of an isolated spiral galaxy in the local universe presenting asymmetric features that we will argue are most likely due 
to cold gas accretion.  This galaxy is drawn from  GASP\footnote{\url{http://web.oapd.inaf.it/gasp/index.html}} (GAs Stripping Phenomena in galaxies with MUSE), an ongoing ESO Large programme granted 120 hours of observing time with the integral-field spectrograph MUSE mounted at the VLT aimed at characterizing where, how and why gas can get removed from galaxies. A complete description of the survey strategy, data reduction and analysis procedures is presented in \cite[][Paper I]{Poggianti2017a}. Briefly, GASP targets 94  ``stripping-candidate galaxies'' in a wide range of environments (from clusters to the general field) selected from a catalog built on a systematic search for galaxies with optical signatures of unilateral debris or tails. Such signatures are reminiscent of gas stripping processes or other processes leading to extraplanar star formation in outgoing gas
\citep{Poggianti2016}.  Targets are selected from WINGS \citep{Fasano2006} and OMEGAWINGS \citep{Gullieuszik2015} surveys and from the Padova Millennium Galaxy and Group Catalogue \citep[PM2GC,][]{Calvi2011}. As all the other targets of GASP, the galaxy we discuss in this paper was selected for presenting a B-band morphological asymmetry suggestive of unilateral debris. 

Throughout all the papers of the GASP series, we adopt a \cite{Chabrier2003} initial mass function (IMF) in the mass range 0.1-100 M$_{\odot}$. The cosmological constants assumed are $\Omega_m=0.3$, $\Omega_{\Lambda}=0.7$ and H$_0=70$ km s$^{-1}$ Mpc$^{-1}$. This gives a scale of 0.912 kpc/$^{\prime\prime}$ at the redshift of P11695, which is $z = 0.04648$.

\section{The target}\label{sec:data}

\begin{figure}
\centering
\includegraphics[scale=0.45]{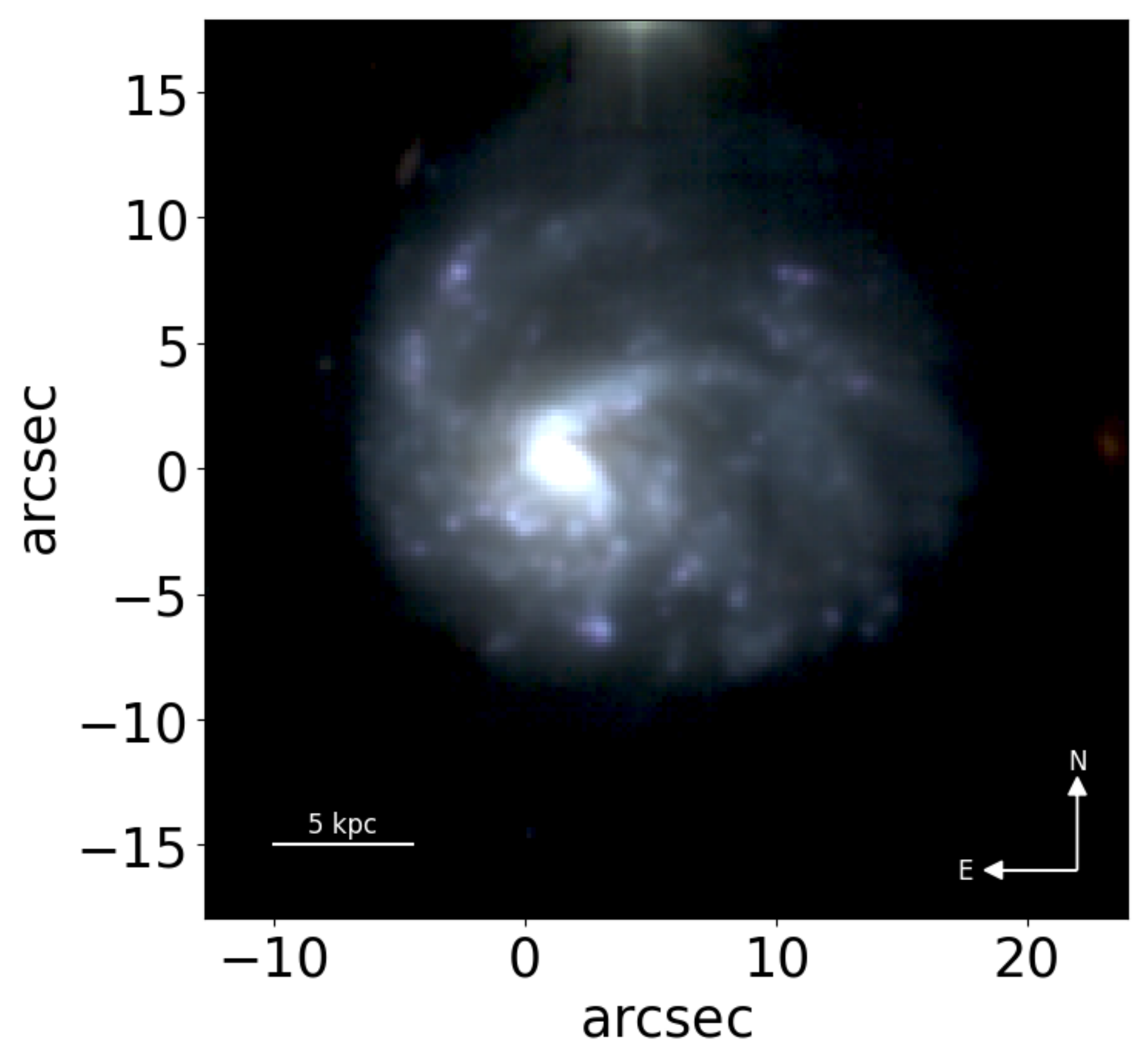}
\caption{RGB image of P11695. The reconstructed  $g$, $r$, $i$ filters from the MUSE cube have been used. At the galaxy redshift, 1$^{\prime\prime}$ = 0.912 kpc, see scale.  North is up, and east is left.  \label{fig:rgb_image} }
\end{figure}

Focus of this analysis is \pp (RA: 10:46:14.78, DEC: +00:03:01.5), a galaxy at z= 0.04648 drawn from the Millennium Galaxy Catalog \citep{Liske2003,Driver2005}.
Figure \ref{fig:rgb_image} shows a color composite image obtained combining the reconstructed $g$, $r$ and $i$ filters from the MUSE datacube. It appears evident that \pp is an asymmetric spiral galaxy possessing trails of bright knots across 
the disk. The galaxy seems also to be characterized by a small U-shaped warp.

\subsection{Observations}

Following the GASP strategy, \pp was observed in
service mode with the MUSE spectrograph, mounted at the Nasmyth focus of the UT4 VLT, at Cerro Paranal in Chile. It was observed on 9/10-Jan-2016, with photometric conditions; the seeing at 650nm (measured on telescope guide star)
remained below 0$\farcs$7 during the whole observing block. A total of four 675 second exposures
were taken with the Wide Field Mode.

MUSE is composed by 24 IFU modules and each of them is equipped with a 4k$\times$4k CCD. The spectral coverage  spans the wavelength range between
4500\AA{} and 9300\AA{}, with a sampling of 1.25 \AA{}/pixel and a spectral resolution of  $\sim$2.6\AA{}. The $1^\prime \times 1^\prime$ field of view is sampled at $\sim0 \farcs 2$/pixel; each datacube therefore consists
of $\sim$10$^5$ spectra.

\subsection{Data reduction}
The data reduction process for all galaxies in the GASP survey is presented in  \citetalias{Poggianti2017a}.  
Briefly, raw data were reduced using the latest ESO MUSE pipeline available at the time of observations (v1.2.1). 

As the data have sufficient sky coverage, the sky was modeled directly and subtracted from individual frames using the 20\% pixels with the lowest counts. After wavelength calibration using arc lamp exposures, the final wavelength adjustments were made using sky emission lines. The final, flux-calibrated data cube was generated by lining up and combining the individual frames using sources in the white light images and consists of 329$\times$329 spectra with radial velocities corrected to the barycenter of the Solar System.
The FWHM of point-like sources on the image obtained by convolving the final MUSE datacube with the V-band filter is 0$\farcs$8 (4 pixels).

\section{Analysis}
 
\begin{figure}
\centering
\includegraphics[scale=0.45]{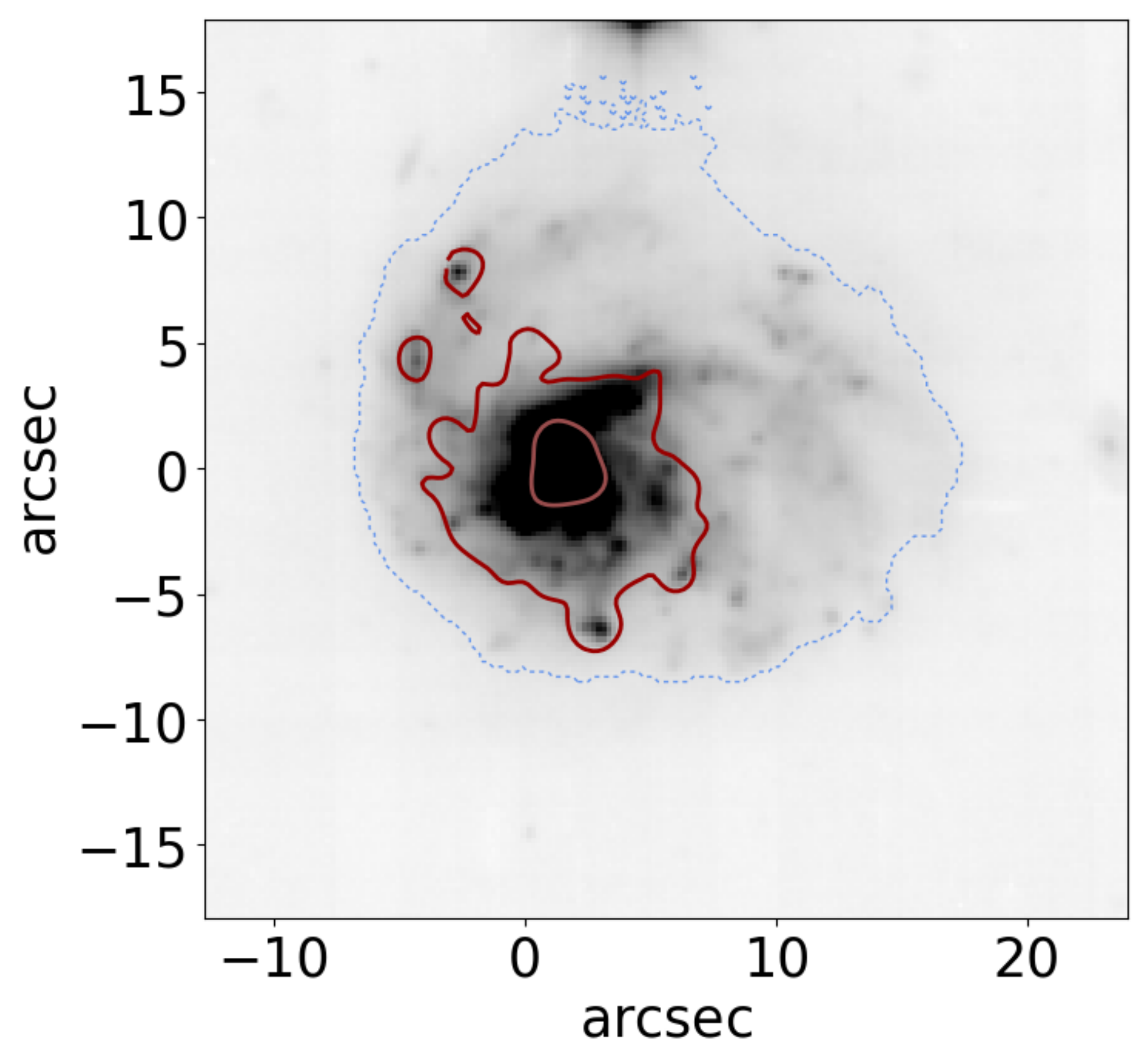}
\caption{MUSE white-light image of P11695. Red solid contours are the distribution of the oldest stellar population (see Fig.\ref{fig:SFH_maps}), which from now on we call ``original body''. The light blue dashed contour shows the ``main body''.  \label{fig:white} }
\end{figure}

\begin{figure*}
\centering
\includegraphics[scale=0.45,clip, trim=5 0 140 0]{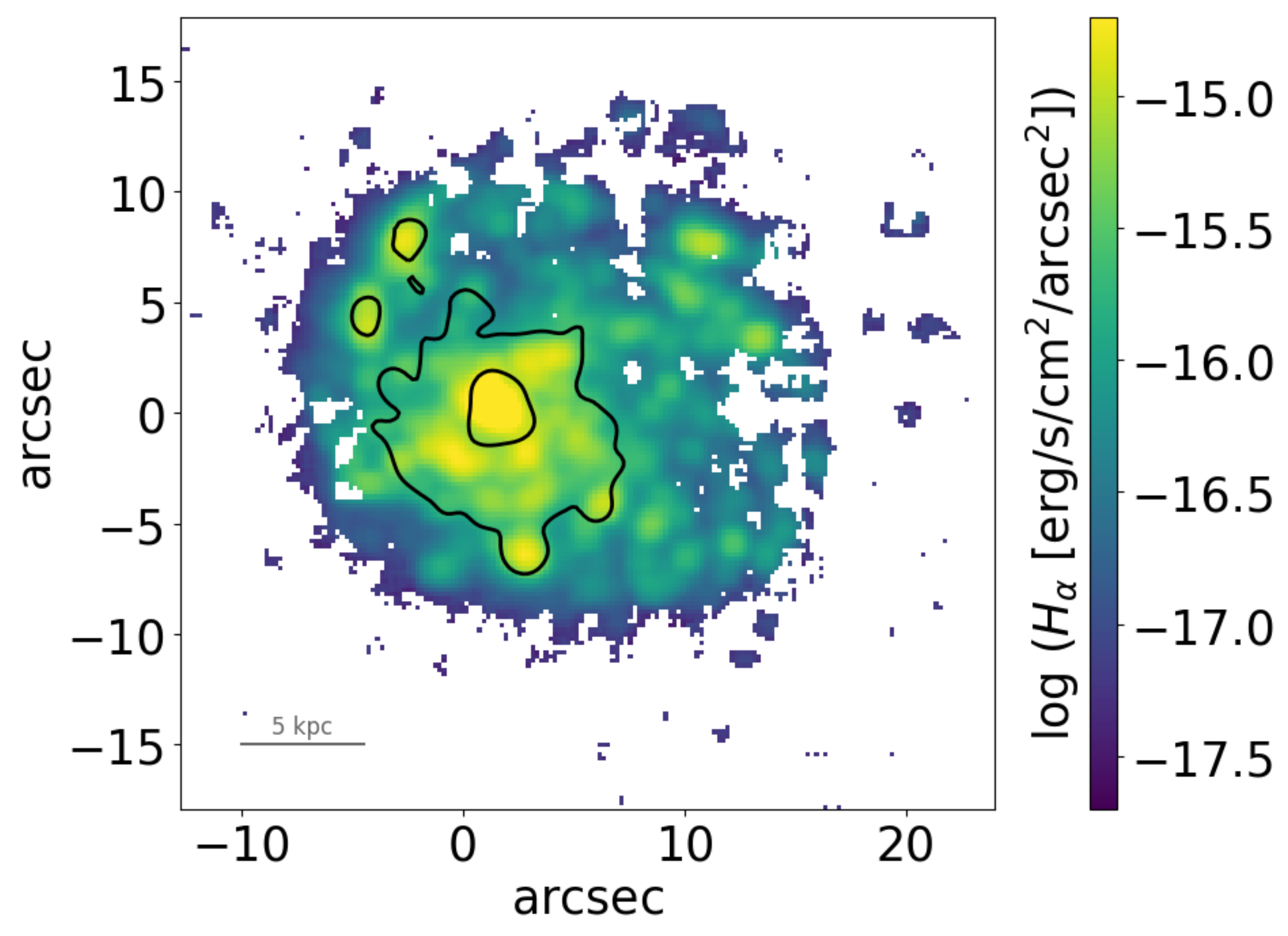}
\includegraphics[scale=0.45,clip, trim=95 0 5 0]{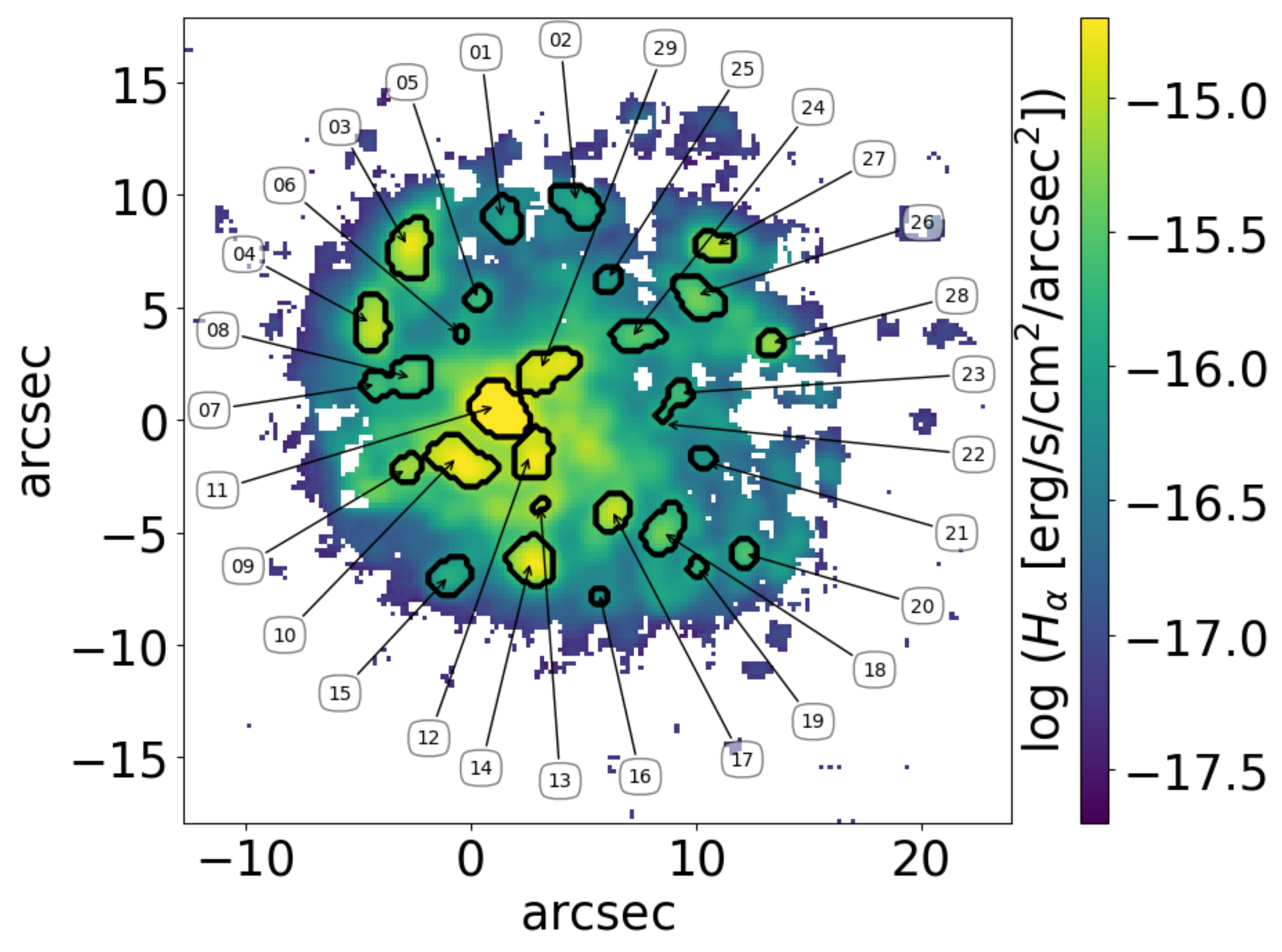}
\caption{MUSE map (median filtered 5$\times$5 pixels) for \Ha S/N$> 4$, uncorrected for stellar absorption and intrinsic dust extinction, but corrected for Galactic extinction. {\it Left.} Contours represent the original body (see Fig.\ref{fig:SFH_maps}). {\it Right.} Contours identify 29 bright \Ha knots, which are also identified by a number (see text for details). \label{fig:Ha}}
\end{figure*}

Figure \ref{fig:white} shows the white-light image from MUSE (4750-9350 \AA{}). From now on, we will use the contours representing the distribution of the oldest stellar population (see Sect. \ref{sec:sinopsis} and Fig. \ref{fig:SFH_maps}) to define the ``original body'' of the galaxy. 
Instead, we will call ``main body'' of the galaxy the region containing the spaxels whose near-\Ha continuum flux is $\sim 1\sigma$ above the background level \citepalias[see also][]{Poggianti2017a}.

The lopsided shape of the galaxy clearly stands out.  Traces of debris to the North and West of the main body and bright knots distributed across the galaxy appear evident. 

The presence of the knots becomes even more striking in  the MUSE \Ha map, shown in Fig. \ref{fig:Ha}. The standard  procedure developed within the GASP project to identify these \Ha knots (see \citetalias{Poggianti2017a} for details)  detected  101 knots, less than 10 of which are outside the main body.  
However, this automatic procedure assumes all the knots have  a round shape, but, as it is evident from Fig.  \ref{fig:Ha}, this is not the case for \pp. We therefore developed another procedure that selects all the regions with a $\rm \log (H\alpha[erg/s/cm^2/arcesc^2])$ flux larger than the (arbitrary value of) $-16.2$. 
Many other fainter knots are visible, but, to be conservative, we choose to focus only on the brightest ones. The right panel of  Fig. \ref{fig:Ha} shows the 29 bright knots we detected with this approach. We note that while many other stripping candidates in the GASP sample are characterized by a conspicuous number of knots, usually these are located mainly in the stripping tails and have a round shape (\citetalias{Poggianti2017a}; \citealt{Gullieuszik2017}, Paper IV). In contrast, almost all the knots in \pp are located on the galaxy disk and they clearly show elongated shapes. These peculiarities induce us to pay particular attention to the properties of the knots, and in what follows we will characterize them separately. 

In the left panel of Fig.  \ref{fig:Ha} we overplot the contours of the original body to the \Ha flux map. The latter is much more extended than the older stellar population, especially on the West side of the galaxy, showing how the galaxy grew in size over time.

\subsection{Dust}

\begin{figure*}
\centering
\includegraphics[scale=0.36]{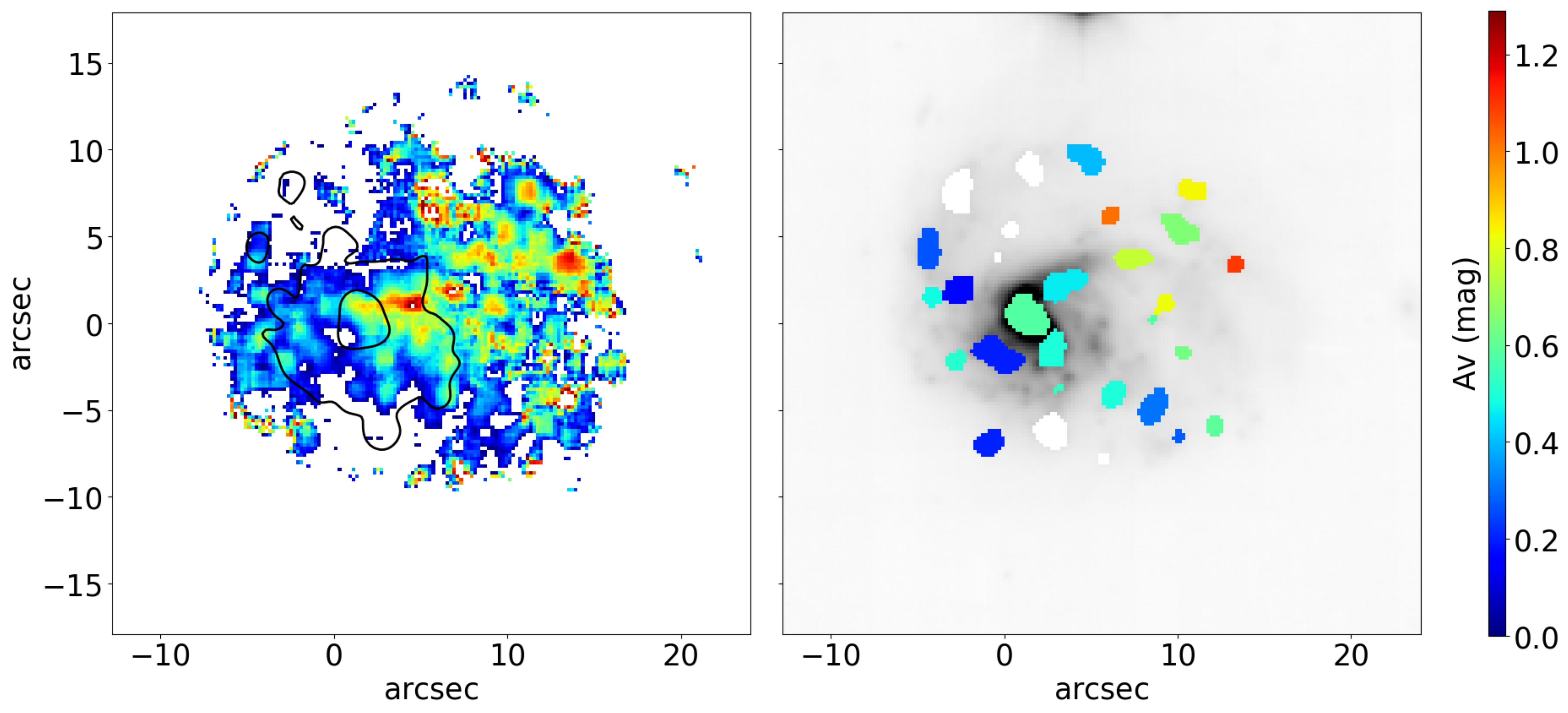}
\caption{$A_V$ map. Only spaxels with a S/N(\Ha)> 3  are shown. {\it Left.} Contours represent the original body (see Fig.\ref{fig:SFH_maps}). {\it Right.} Only regions corresponding to the 29  \Ha knots identified in Fig.\ref{fig:Ha} are color coded according to the median $A_V$ of each knot. \Ha knots with $>50\%$ of the spaxels without determination of the H$\beta$ are colored in white. In the background, the white-light image of the galaxy (Fig.\ref{fig:white}) is shown, for reference. \label{fig:av}}
\end{figure*}

The $\rm H\alpha$ fluxes  presented in Fig. \ref{fig:Ha}
were corrected for extinction by dust internal to \pp.
The map of the dust extinction  A$_V$ is obtained  from the 
absorption-corrected 
Balmer decrement in each spaxel. As described in \citetalias{Poggianti2017a}, we assumed an intrinsic \Ha/H$\beta$ ratio equal to 2.86 and adopt the \cite{Cardelli1989} extinction law. 
The A$_V$ map has been calculated only for spaxels where the S/N on the H$\alpha$ and H$\beta$ lines is larger than 3 and the ratio of the two lines is larger than the assumed 2.86 value for the Balmer decrement. 

The $A_V$ map (Fig. \ref{fig:av}) shows that overall \pp is characterized by low values of extinction, almost always $<1$mag. The North-West side of the galaxy has a systematically larger extinction that the South-East side. Note, however, that the in the East side of the galaxy there are many regions where  $A_V$ is not measured.   In these regions indeed H$\beta$ might be undetected  because of obscuration by dust.

To compute the dust extinction of the \Ha knots, we assign the median value within each  knot to all the spaxels in that knot, therefore spaxels with no measurement in the left hand panel of Fig. \ref{fig:av} might have an associated measurement in the right hand panel. The $A_V$ estimate in the knots can be  therefore considered as a lower limit of the true values. However, we have compared these values to those obtained by integrating over the spectra in each knot, finding  good agreement.
We find that  not all the knots are highly extincted, as we would expect in strongly star forming regions. Overall, there does not seem to be a strong correlation  between the  position of the knots and the distribution of the dust.

\subsection{Gas ionization mechanism}
\begin{figure}
\centering
\includegraphics[scale=0.44]{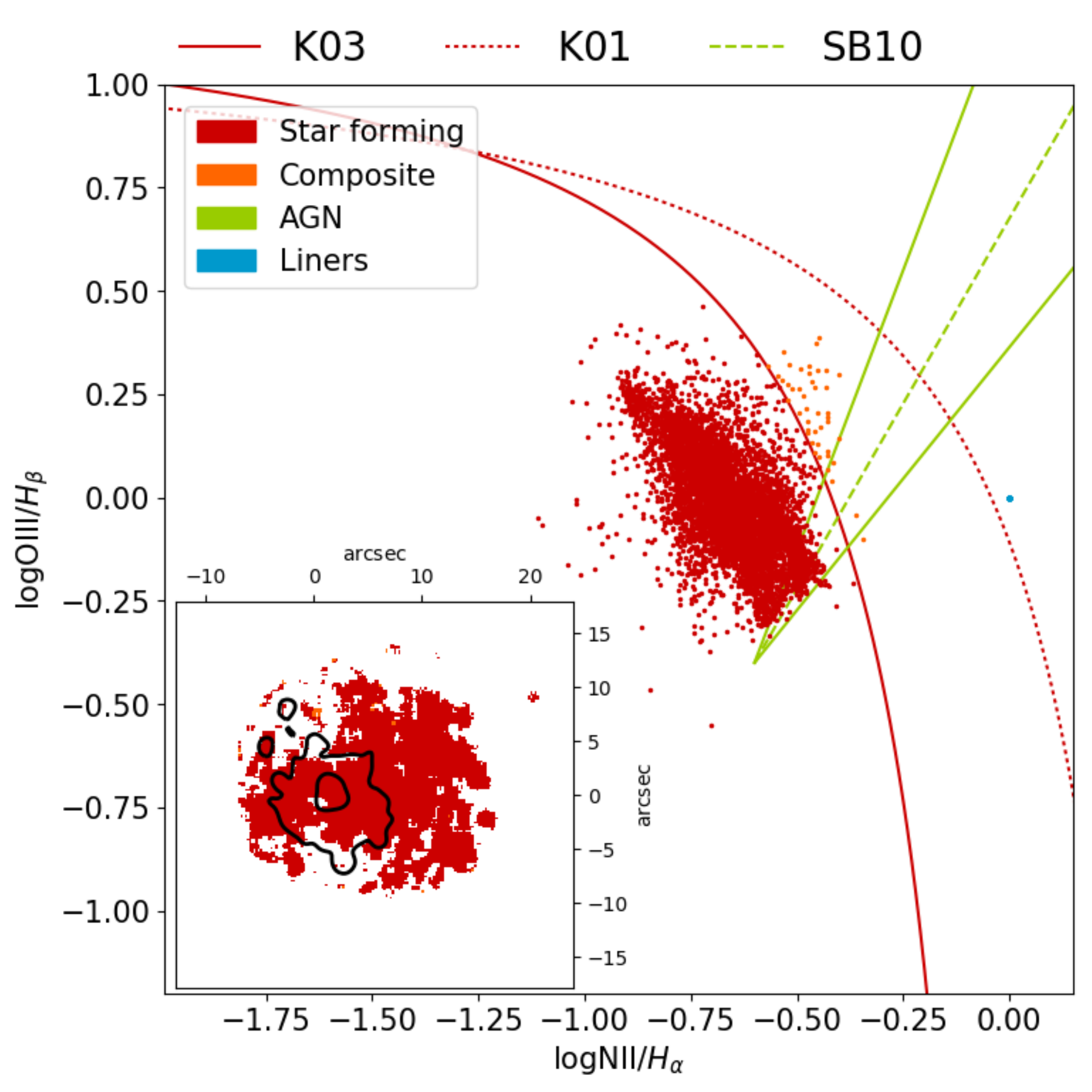}
\caption{BPT line-ratio  diagram for [OIII]5007/H$\beta$ vs [NII]6583/\Ha. Lines  are from \citet[][K03]{Kauffmann2003}, \citet[][K01]{Kewley2001} and \citet[][SB10]{Sharp2010} to separate Star-forming, Composite, AGN and LINERS. In the inset the BPT line-ratio map is shown. Only spaxels with a $S/N> 3$ in all the emission lines involved are shown.
 \label{fig:AGN}}
\end{figure}

The next step is to characterize the ionizing sources as a function of the position,
and detect the possible presence of an active galactic
nucleus (AGN). We therefore inspect the diagnostic diagrams \citep[e.g.,][]{Kewley2006} based on the emission
lines located within our cube observed range (i.e. H$\beta$, [OIII] 5007 \AA{}, [OI] 6300 \AA{}, \Ha, [NII] 6583 \AA{}, and [SII] 6716+6731 \AA{}). The lines' intensities were measured after subtraction of the continuum, exploiting the pure stellar emission best fit model provided by our spectral fitting code {\sc sinopsis} \citep[][Paper III, see Sect. \ref{sec:sinopsis}]{Fritz2017}, to take into account any possible contamination from stellar photospheric absorption. Only spaxels with a $S/N> 3$ in all the emission lines involved were considered.

All the diagnostic tools  
are concordant that  the emission-line ratios are  consistent with gas being photoionized by young stars \citep[``Star-forming'' according to][]{Kauffmann2003, Kewley2006}. As an example, Fig. \ref{fig:AGN} shows the [OIII]5007/H$\beta$ vs [NII]6583/\Ha.  

\cite{Veron2003} classified \pp (identified as SDSS J10462+0003) as Seyfert 1, but this catalog  is obsoleted by  \cite{Veron2006}, which does not detect any sign of AGN activity, in agreement to our results.

We also characterize the main source of ionization for the \Ha knots and, according to BPT diagrams, they are all powered by star formation (plot not shown).

\subsection{The integrated and spatially resolved Star Formation Rate} \label{sec:SFR}

\begin{figure}
\centering
\includegraphics[scale=0.31]{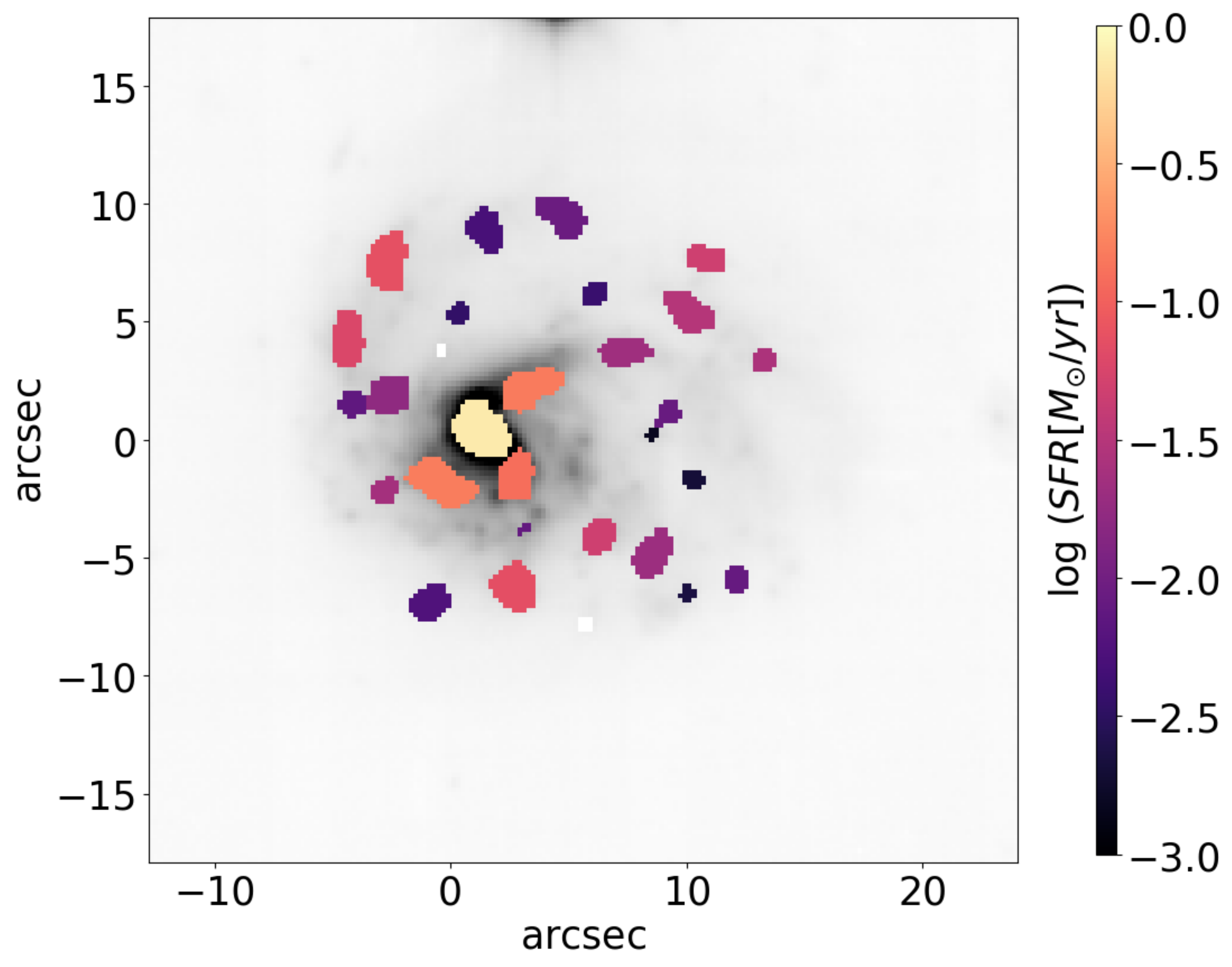}
\caption{Total SFR produced in each of the 29  \Ha knots identified in Fig.\ref{fig:Ha}. In the background, the white-light image of the galaxy (Fig.\ref{fig:white}) is shown, for reference. \label{fig:SFR}}
\end{figure}

Having assessed that the ionization source in \pp is mostly photoionization by young stars, we can now measure the total ongoing Star Formation Rate (SFR), obtained
from the dust- and absorption-corrected \Ha luminosity   adopting the \cite{Kennicutt1998a}'s relation for a \cite{Chabrier2003} IMF: $\rm SFR = 4.6\times  10^{-42} L_{H\alpha}$.
Integrating the spectrum over the galaxy main body, for the spaxels with a S/N(\Ha)$> 3$, we get a value of SFR=3.27 $\rm{M_\odot \, yr^{-1}}$. The total SFR in the \Ha knots is 1.70 $\rm{M_\odot\, yr^{-1}}$: half of the  current SFR is therefore taking place in these HII regions.  This value can be considered a lower limit to the total SFR in the knots, indeed the contribution of the \Ha flux escaped from the knots is not taken into account.

Figure \ref{fig:SFR} shows the spatial distribution of the SFR in the \Ha knots. The SFR in all the spaxels included in each knot is summed up. The  central knot has a very high value of SFR, while all the other regions are producing stars at a much lower rate ($<1:3$). 

\subsection{The stellar and gas kinematics}

\begin{figure*}
\centering
\includegraphics[scale=0.38]{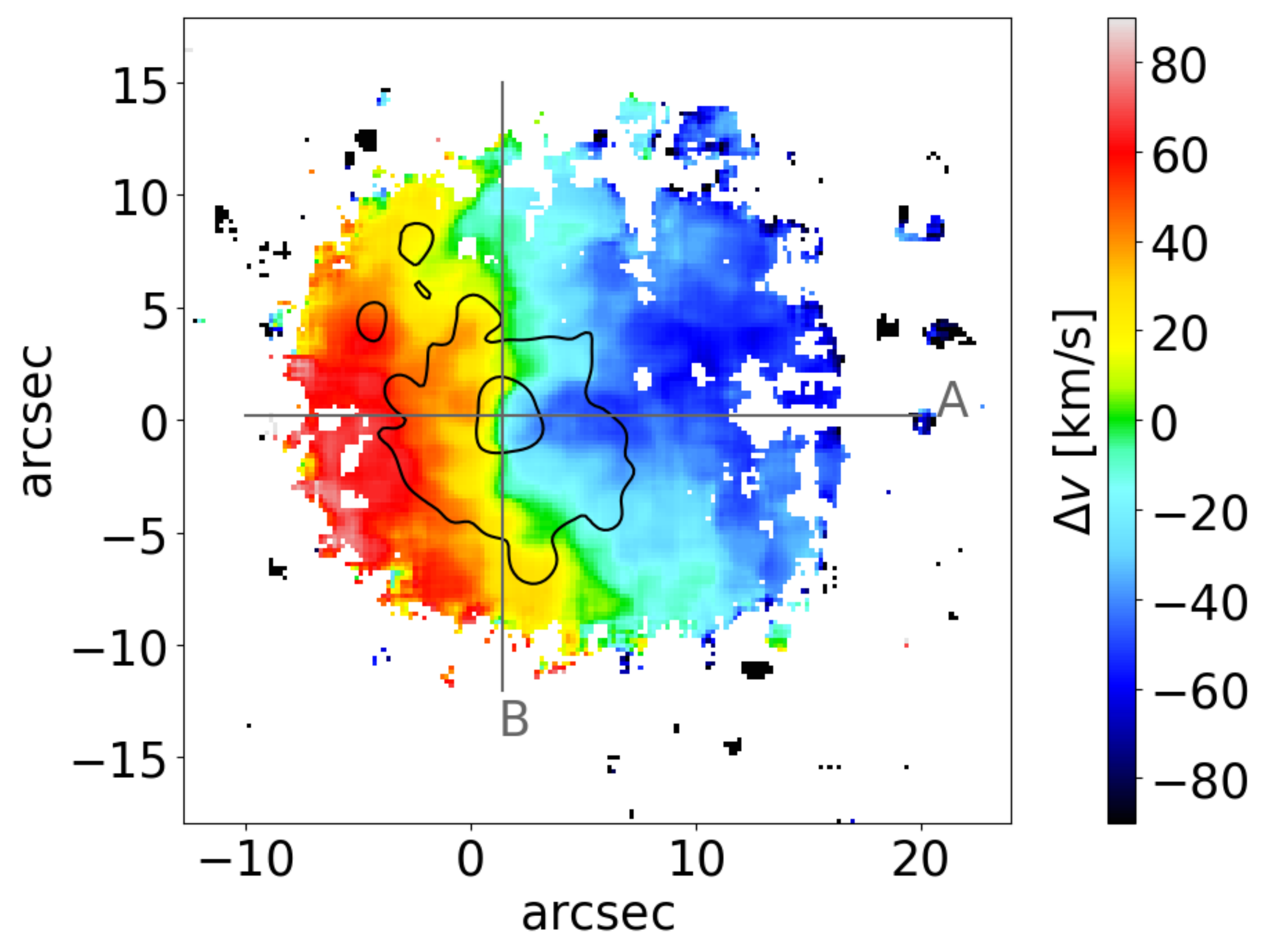}
\includegraphics[scale=0.38]{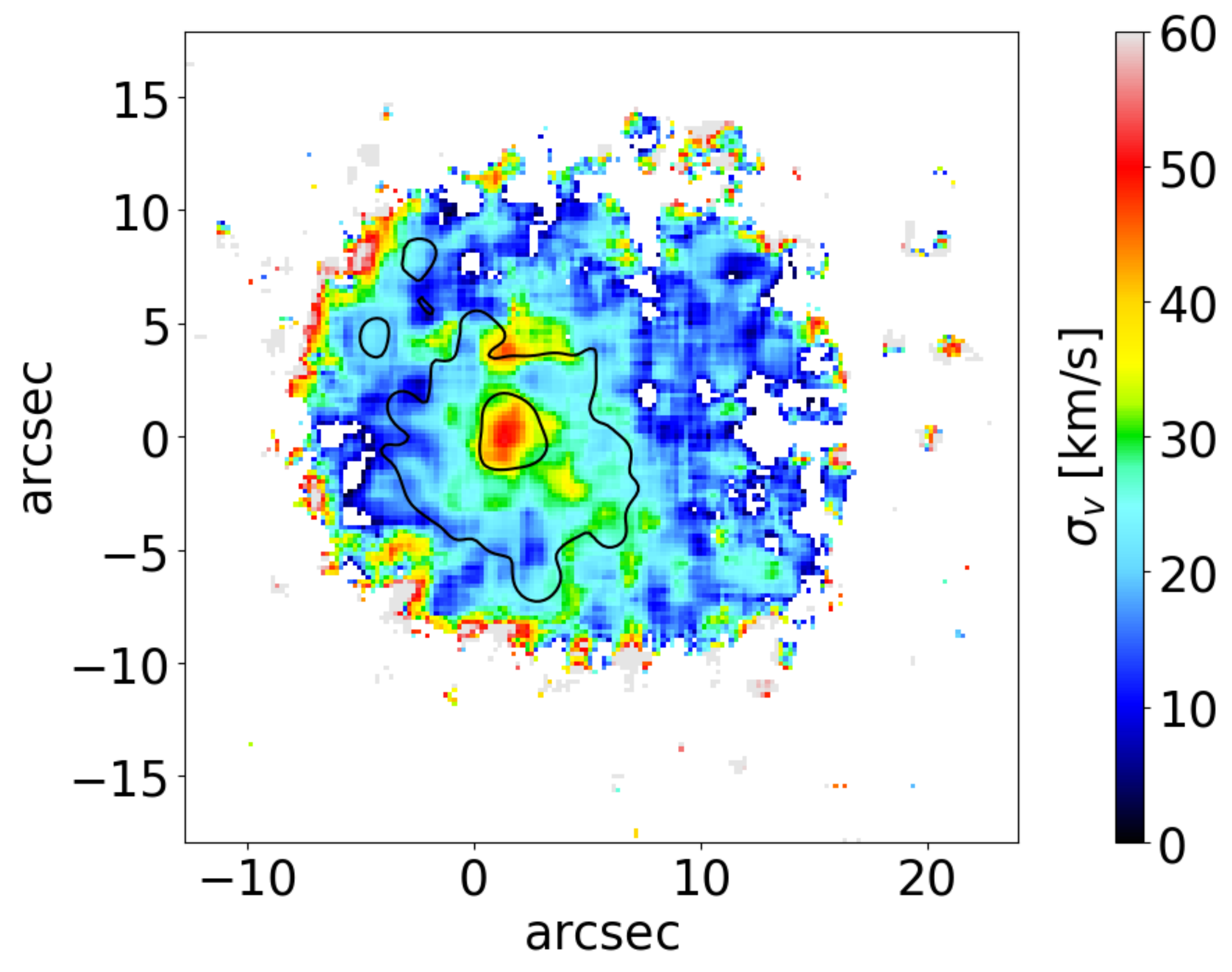}
\includegraphics[scale=0.38]{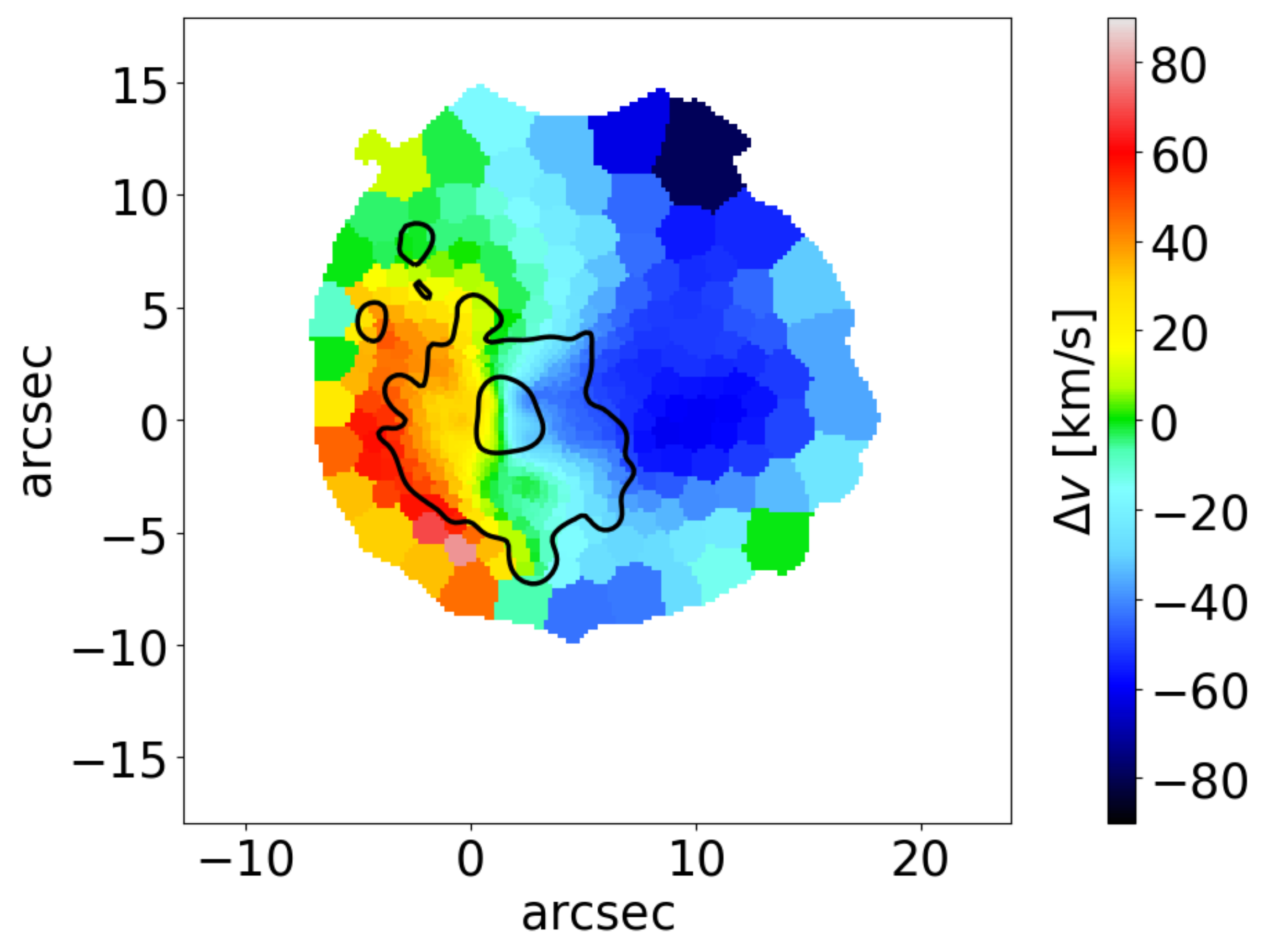}
\includegraphics[scale=0.38]{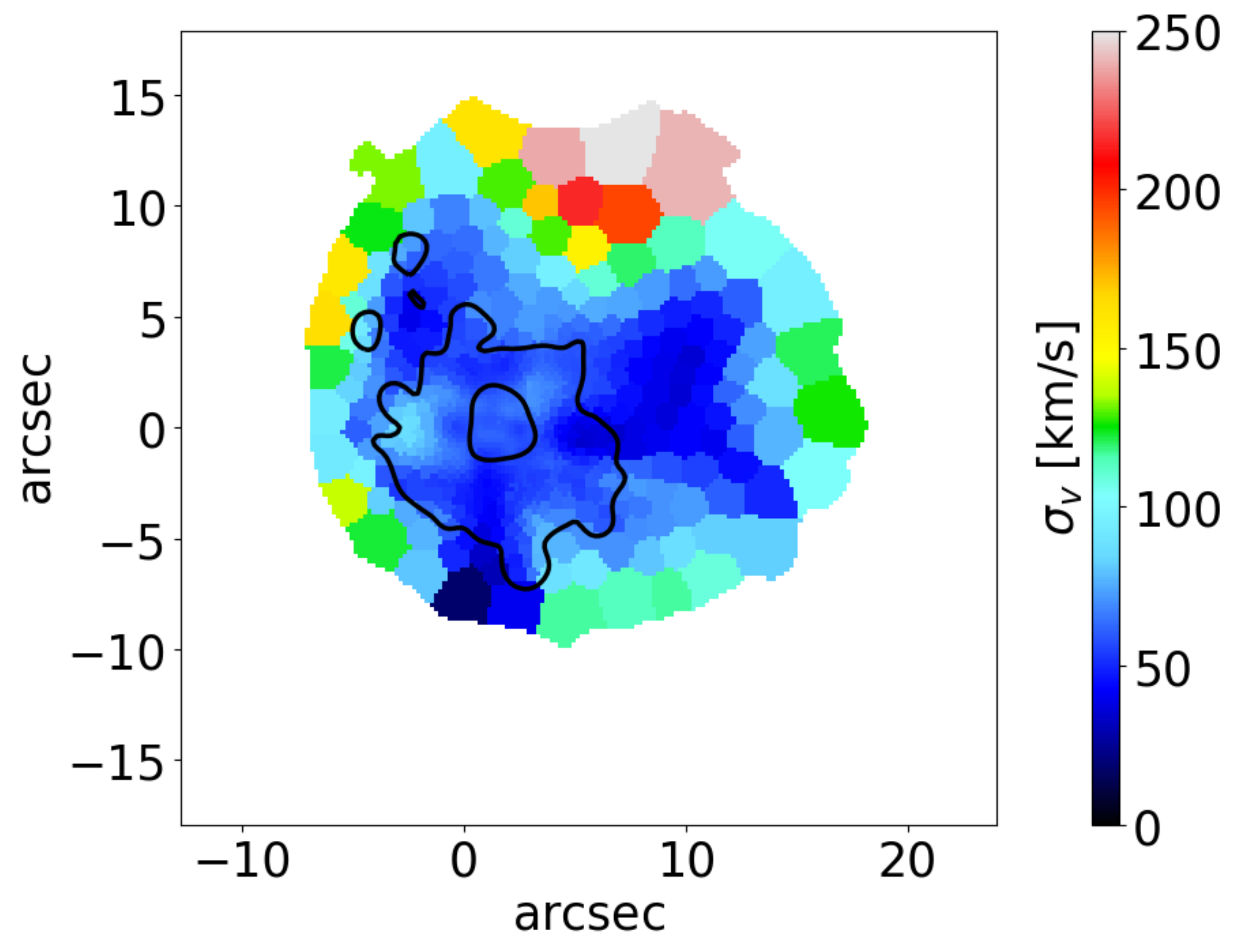}
\caption{{\it Top.} \Ha velocity (left) and velocity dispersion (right) map for 5$\times$5 spaxels with S/N \Ha > 4. {\it Bottom.}  Stellar velocity (left) and velocity dispersion (right) map for Voronoi bins with S/N> 15. Contours represent the original body (see Fig.\ref{fig:SFH_maps}). In the \Ha velocity map the axis along which profiles are extracted are shown (A and B).  \label{fig:vels_gas}}
\end{figure*}

\begin{figure}
\centering
\includegraphics[scale=0.33]{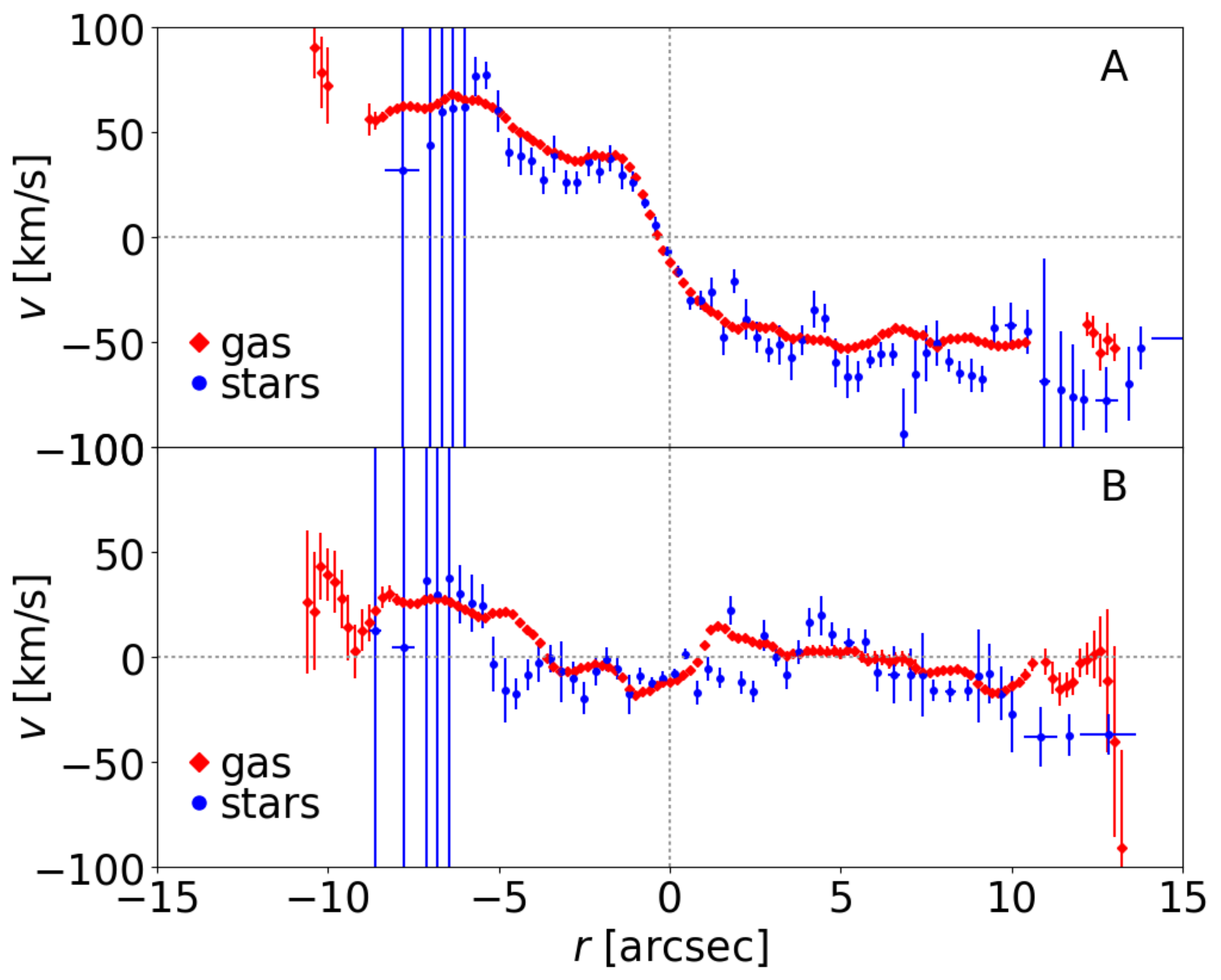}
\caption{Velocity profiles along the two slits shown in Fig.\ref{fig:vels_gas} (A, B), as indicated in the labels, both for the gas and the stars. Distances are deprojected considering the galaxy inclination and position angle. Values and vertical errors s are weighted by the formal errors, for the stellar velocity, horizontal errors indicate the width of the Voronoi bins.  \label{fig:prof_vel}}
\end{figure}

We now characterize separately the kinematics of the gas and stellar components (Fig. \ref{fig:vels_gas}). 

The kinematic properties of the gas were  inferred by the analysis of the characteristics of emission lines, especially \Ha, by exploiting the {\sc kubeviz}  \citep{Fossati2016} code. 
In the MUSE wavelength range the typical spectral dispersion of 1.25 \AA{} pixel$^{-1}$ translates to a velocity scale of 25 km s$^{-1}$ pixel$^{-1}$. The average FWHM
resolution is 2.51 \AA{}, equivalently to 110 km s$^{-1}$. Details on the methods can be found in \citetalias{Poggianti2017a}.

The stellar kinematic was instead derived from the analysis of the characteristics of absorption lines, using the pPXF software \citep{Cappellari2012}, which works
in Voronoi binned regions of given S/N \citep[15 in this case; see][]{Cappellari2012_v}. The value of the stellar radial velocity was further smoothed using the two-dimensional local regression techniques (LOESS) as implemented in the Python code developed by M. Cappellari.\footnote{\url{http://www-astro.physics.ox.ac.uk/~mxc/software}}

The  velocity fields of the two components are co-rotating around the same axis,  approximately the North-South direction (slit B in Fig. \ref{fig:vels_gas}),  and span a similar velocity range ($-90<v (\rm km \, s^{-1}) <90$).\footnote{We applied  a shift in velocity of 25 $\rm km \, s^{-1}$ to the gas to force the velocities of the two components to be the same in the galaxy center.} In both cases, the West side 
is characterized by negative velocities, indicating that side is approaching.  For the gas uncertainties are in most of the spaxels <2.5 $\rm km \, s^{-1}$, except in the very external regions. Uncertainties on the stellar motion are the formal errors of the fit calculated using the original noise spectrum datacube and have been normalized by the $\chi^2$ of the fit. These errors are systematically larger ($\sim20-30 \rm km \, s^{-1}$) than those for the gas, reflecting the quality of the fit, but still much smaller than the measured velocities, reassuring us about the robustness of our results. 

In the gas, the locus of negative velocities shows a bending in the external regions, with the convexity pointing toward East. 

The velocity dispersion maps highlight some differences between the two components.  In the original body,  the $\sigma_{gas}$ goes from $\sim 15$ to 50 $\rm km \, s^{-1}$,  while $\sigma_{stars}$  ranges from 50 to $\sim 75$ $\rm km \, s^{-1}$ and does not increase toward the center.
In the outskirts, the velocity dispersion of the gas is always relatively  low ($\sim 10-20$ $\rm km \, s^{-1}$), indicative of a dynamically cold medium, except on the South and East edges of the galaxy, but there  the low S/N ratio of these spaxels prevents us from drawing solid conclusions. In contrast,  the motion of the stars is much more random, as indicated by the stellar $\sigma$ that reaches peaks of  $\sigma >250$ $\rm km \, s^{-1}$, in the Northern regions. Nominal errors on the dispersions are similar to uncertainties on the velocities, both for the gas and the stars.

To better contrast the  gas and stellar kinematics,  
we extract the velocity profiles (Fig. \ref{fig:prof_vel}) along the rotation axis and its perpendicular direction (B and A in Fig. \ref{fig:vels_gas}, respectively). 
We deproject the radius considering the position angle and inclination (the inferred position angle is 40${\degree}$, the inferred inclination is 42${\degree}$). We compute the average value of the spaxels entering the slit at each distance, weighted for the corresponding errors.
Along the slit A, the gas presents a quite regular rotation. In the central part of the galaxy ($\pm3^{\prime\prime}$) the variation of the velocity is quite steep, and then it flattens out and the curves are almost flat. The stars follow the same trend, even though the curve is more noisy.  In contrast,  along the slit B, for both components the curve is always quite flat, even though with many local variations. This was expected since the slit A is approximately 90${\degree}$ from the rotation axis.

\subsection{The spatially resolved gas  properties}

\begin{figure*}[!t]
\centering
\includegraphics[scale=0.36]{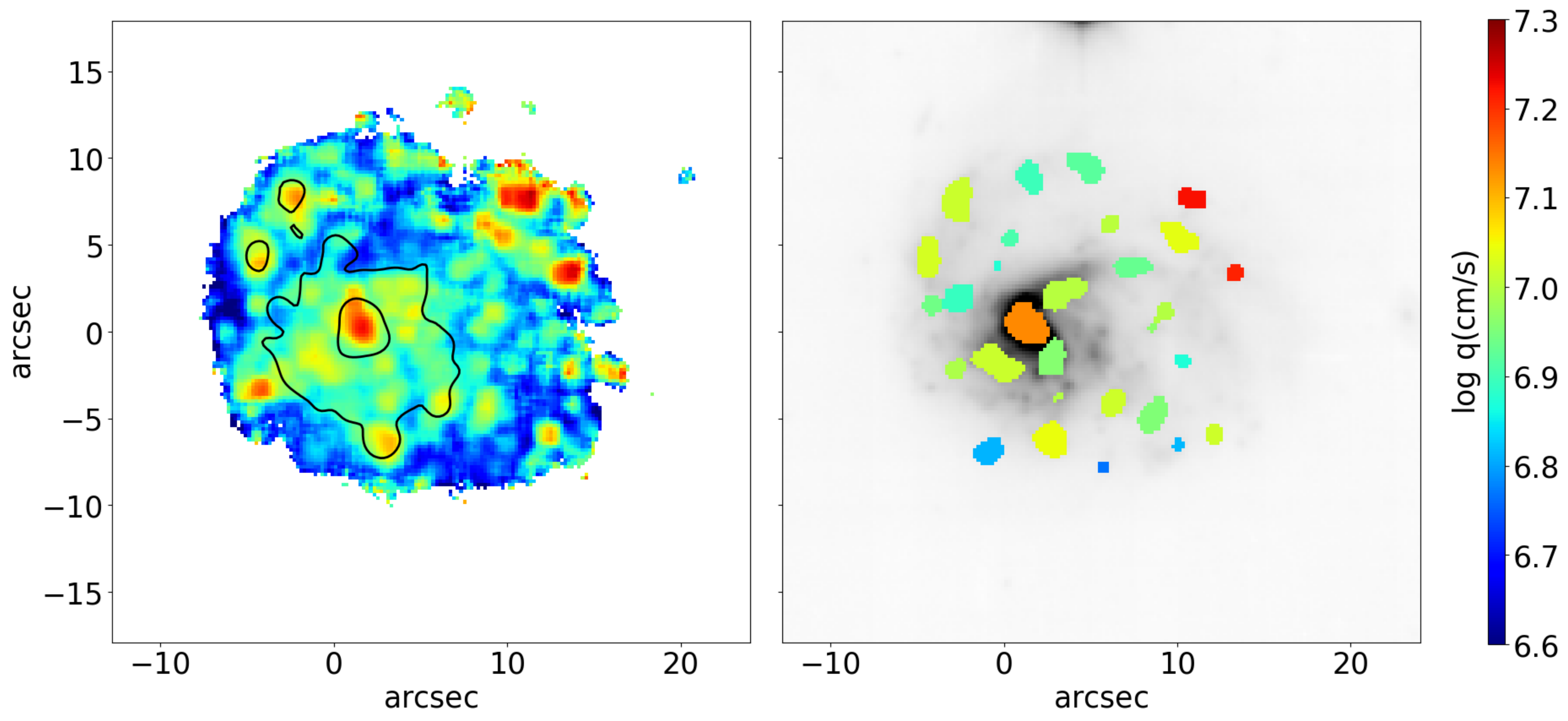}
\caption{Ionization parameter (q) map. {\it Left.} Contours represent the original body (see Fig.\ref{fig:SFH_maps}). {\it Right.} Only regions corresponding to the 29  \Ha knots identified in Fig.\ref{fig:Ha} are color coded according to the median q of each knots. In the background, the white-light image of the galaxy (Fig.\ref{fig:white}) is shown, for reference. \label{fig:q}}
\end{figure*}

\begin{figure*}
\centering
\includegraphics[scale=0.36]{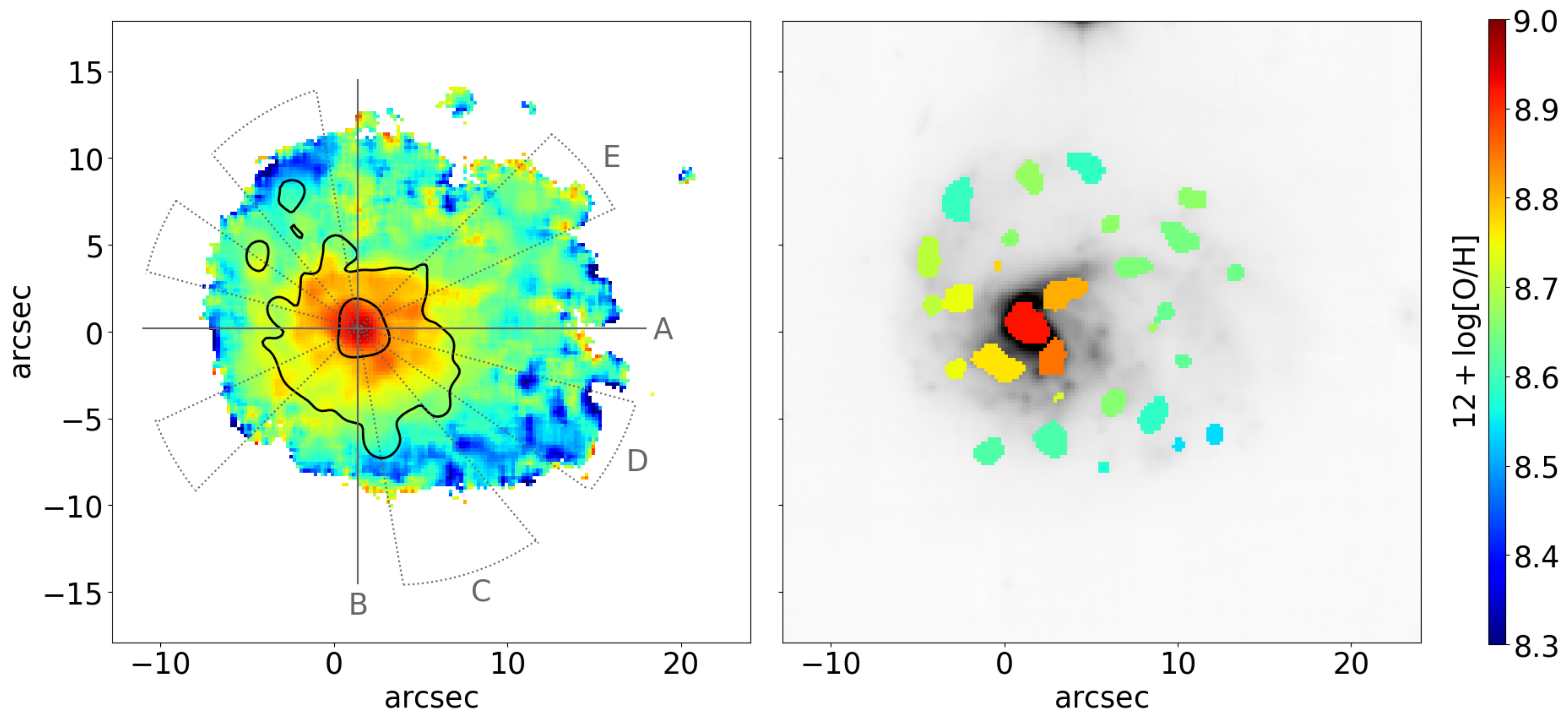}
\caption{Gas metallicity map. {\it Left.} Contours represent the original body (see Fig.\ref{fig:SFH_maps}). Slits (A, B) and conic apertures (C, D, E) used to extract metallicity gradients are also shown. {\it Right.} Only regions corresponding to the 29  \Ha knots identified in Fig.\ref{fig:Ha} are color coded according to the median metallicity of the knots. In the background, the white-light image of the galaxy (Fig.\ref{fig:white}) is shown, for reference.\label{fig:z}}
\end{figure*}

\begin{figure}
\centering
\includegraphics[scale=0.25]{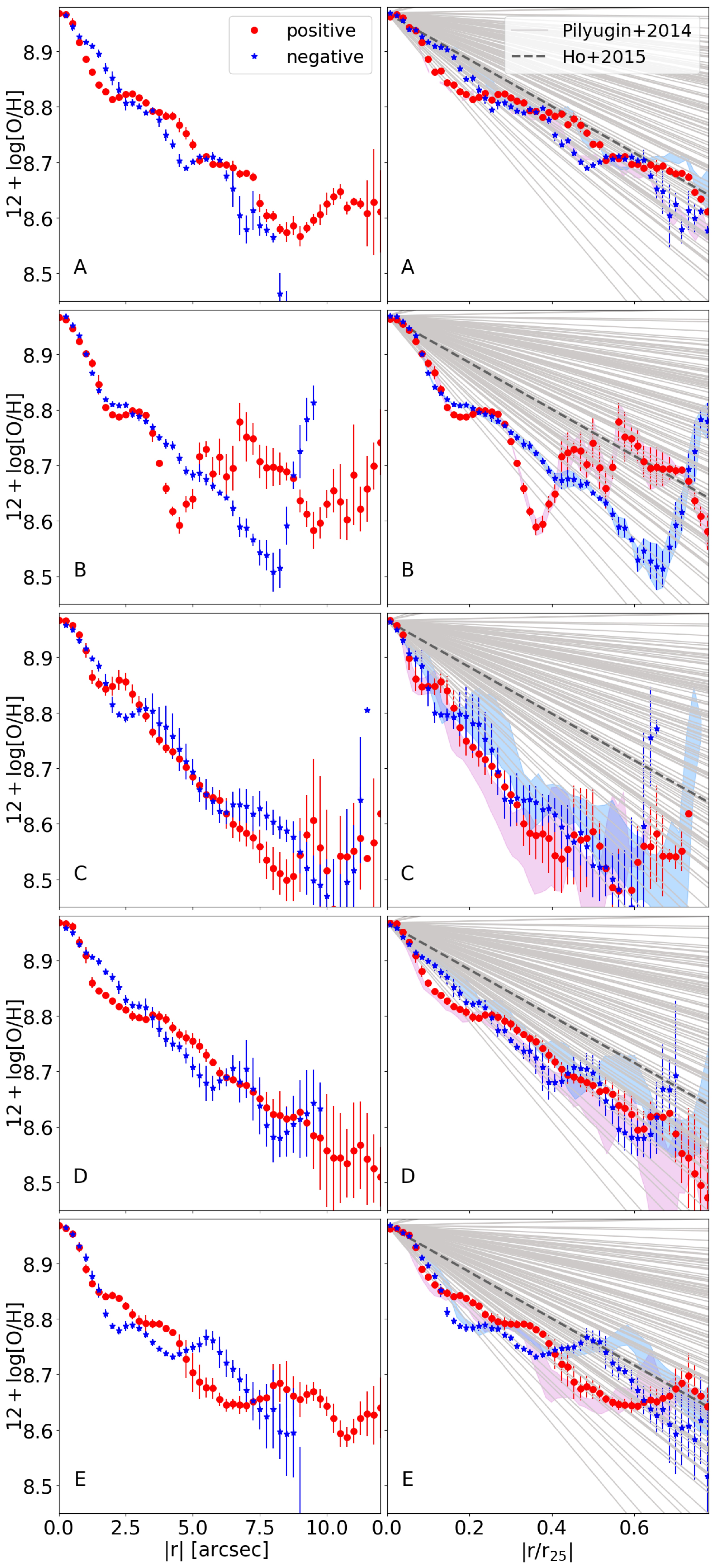}
\caption{Gas metallicity gradients extracted along the directions (A, B, C, D, E) shown in Fig.\ref{fig:z}. Errors represent the standard deviation ($\sigma$) computed using the values falling into the apertures. Red points correspond to the side of the slits Northern and/or Western the center (positive arcsecs), blue points correspond to the  side of the slits Southern and/or Eastern the center (negative arcsecs). {\it Left.} Profiles in arcsec, not deprojected. {\it Right.} Profiles normalized by a scale radius, all distances are deprojected considering the galaxy inclination and position angle. Shaded areas represent the profiles normalized adopting different scale lengths (see text for details).  Grey lines are from \citet{Pilyugin2014,Ho2015}. \label{fig:z_grad}}
\end{figure}

The gas metallicity and ionization parameter for
each star forming spaxel were calculated using the pyqz Python
module7 \citep{Dopita2013} v0.8.2; the 
$12 + \log(O/H)$ and $\log(q)$ values are obtained by interpolating
from a finite set of diagnostic line ratio grids computed
with the MAPPINGS code. We used a modified version
of the code (F. Vogt, priv comm.) to implement the
MAPPING IV grids that are calibrated in the range
$7.39 < 12 + \log(O/H) < 9.39$, which is broader than
the metallicity range covered by MAPPING V grids.
Our results are based on the [NII]6585 / [SII]6717+6731
vs [OIII]/[SII]6717+6731 (see \citetalias{Poggianti2017a}).
As discussed in  detail by \cite{Kewley2008}, the systematic errors introduced by modeling inaccuracies are usually estimated to be $\sim$0.1-0.15 dex, whereas discrepancies of up to 0.2 dex exist among the various calibrations based on photoionization models. 

The ionization parameter is quantified as the ionizing photon flux through a unit area divided by the local number density of hydrogen atoms. The spatial distribution of the ionization parameter for \pp is presented in Fig. \ref{fig:q}. It is generally 
low, ranging from  6.6$<\log(q)<$7.3. This is in agreement with other galaxies in the GASP survey  (\citetalias{Poggianti2017a,Gullieuszik2017}), even though in this case a clear clumpy pattern emerges, while in other galaxies a more smooth distribution is observed. 
\Ha knots tend to have a systematically large value of q: the median value integrated across the entire galaxy is  6.8, while the median value in the \Ha knots is  7.

Figure \ref{fig:z} shows that the spatially resolved metallicity ranges between { 8.3}$<12+\log(O/H)<${9}, with a median value of { 8.6.} The maximum of the metallicity corresponds to the maximum of the \Ha flux. According to the \cite{Tremonti2004} mass-metallicity relation, a galaxy with mass similar to that of \pp should have a metallicity of $12+\log(O/H)\sim8.99$. However, 
 the uncertainties in the absolute
calibration of the metallicity scale (F. Vogt, priv. comm.) prevent us from drawing solid conclusions regarding the low metallicity of \pp. 

The galaxy shows a sharp decrease in its metallicity from the center toward the outskirts. The metallicity distribution is inhomogeneous and it is not characterized by spherical  symmetry. For example, two regions of very low metallicity stand out in the South-West  and North-East side of the galaxy, while the North-West side has systematically higher metallicity values. 

The metallicity in the \Ha knots (right panel of Fig. \ref{fig:z}) follows the  trend of the entire galaxy:  those in the center are more metal-rich, those in the outskirts have lower values of metallicity, especially on the South-West and North side of the galaxy. 

Trends between metallicity and distance from the galaxy center are more clear  in Fig. \ref{fig:z_grad}, where metallicity gradients along different axes are shown. To start, we use the axis defined by the velocity maps (slits A and B in Figs. \ref{fig:vels_gas} and \ref{fig:z}). In order to get more signal in the galaxy outskirts, instead of extracting gradients along a slit, we select all the spaxels within conic apertures $\pm3{\degree}$ wide around the slits and take the median of the metallicity as a function of distance. This choice also allows us to compute errors as standard deviation within the aperture. 
 Along A, the gradients extracted on the two sides (East negative and West positive) of the galaxy are quite similar, except that on the  West side at $1^{\prime\prime}<|r|<2^{\prime\prime}$ there is a steeper decline in the metallicity values with distance than on the other side. Trends are inverted at $|r|\sim 4^{\prime\prime}$, where the East side is more metal poor than the West side.  Along  B, gradients are symmetric around the 0 position up to 5$^{\prime\prime}$, then the { South side shows a constant decline and a sudden increase at $r>7^{\prime\prime}$, while the North side of the galaxy shows some fluctuations.}  

Clearly, these two axes do not capture the largest variation of metallicity in the galaxy. We therefore use three additional axes (C, D, E), which are chosen ad hoc to better characterize the variations across the galaxy. The chosen apertures are shown in the left panel of Fig. \ref{fig:z}, and are 30${\degree}$, 20${\degree}$ and 20${\degree}$  wide, respectively. While gradients along D are similar to those along B, the panels in the  third row of  Fig. \ref{fig:z_grad} shows that the gradients along C are the steepest ones.  The minimum metallicity is $\log(O/H)\sim$ { 8.4}, while along A and B, for example, it is always $\geq$  { 8.5}. 
Finally, the E aperture catches one of the flattest gradients in the galaxy, with metallicity values >{ 8.6} at all distances. 

To understand if these results are in some sense peculiar or are similar to those of other galaxies in the local universe, in the right panels of Fig. \ref{fig:z_grad}  we compare the metallicity gradients of \pp to those tabulated by \cite{Pilyugin2014},  who  investigated the oxygen  abundance distributions across the optical disks of 130 nearby late-type galaxies,  and by \cite{Ho2015}, who measured the average metallicity of $\sim$50 nearby galaxies mainly drawn from the CALIFA \citep{Sanchez2012} survey. For all the samples, $r_{25} $ has been computed as the major axis of the isophote of surface brightness = 25 in B band and radii are deprojected considering the position angle and inclination. As \pp has a very lopsided optical light distribution, a single scale  length might not be representative of the entire galaxy. Therefore, we also compute $r_{25}$ considering separately the West side and the East side of the galaxy. In the right panels of Fig.\ref{fig:z_grad}, while the profiles represented by the symbols are scaled using the average scale length $r_{25}=16\farcs 25$, profiles shown in shaded areas are scaled using the most appropriate scale length: the positive sides of the profiles extracted along A, C, D and E are scaled using  $r_{25}=18\farcs 57$, the corresponding negative sides are scaled using  $r_{25}=13\farcs 93$. For the profiles along the slit B, the average scale length is adopted.  As expected, the slope of the profiles depends on the adopted scale length, but the general conclusions hold. To renormalize the \citeauthor{Pilyugin2014, Ho2015} samples to ours, we assume that the maximum central metallicity coincide in all the samples. This normalization allows us to compare the gradients  regardless of the uncertainties related to the absolute calibrations adopted by the different works. Overall, \pp presents the  steepest gradients. Differences are the largest for the trends extracted along the aperture C, the smallest for trends extracted along E. \pp is therefore at the tail of the metallicity-radius distribution.

\subsection{The integrated and spatially resolved stellar population properties} \label{sec:sinopsis}
The spectral fitting code {\sc sinopsis} \citepalias{Fritz2017} allows us to characterize the spatially resolved properties of the stellar populations. Details can be found in \citetalias{Poggianti2017a,Fritz2017}. Briefly, this code combines different simple stellar populations  spectra to reproduce the observed equivalent widths of the main absorption and emission lines, and the continuum in various  bands. The main outputs of {\sc sinopsis} are maps of stellar mass, average star formation rate and total mass formed in four age bins,  luminosity-weighted and mass-weighted stellar ages. In addition, it also produces a best-fit model datacube for the stellar-only component.

Running {\sc sinopsis} on the galaxy integrated spectrum obtained within the galaxy main body, the total stellar mass measured is $M_\ast=1.17 \times 10^{10} M_\odot$. Considering the total SFR measured in Sec. \ref{sec:SFR},  \pp lays on the upper envelope of the typical SFR-mass relation for star-forming field galaxies  \citep[see][]{Poggianti2016}.

\begin{figure*}
\centering
\includegraphics[scale=0.36]{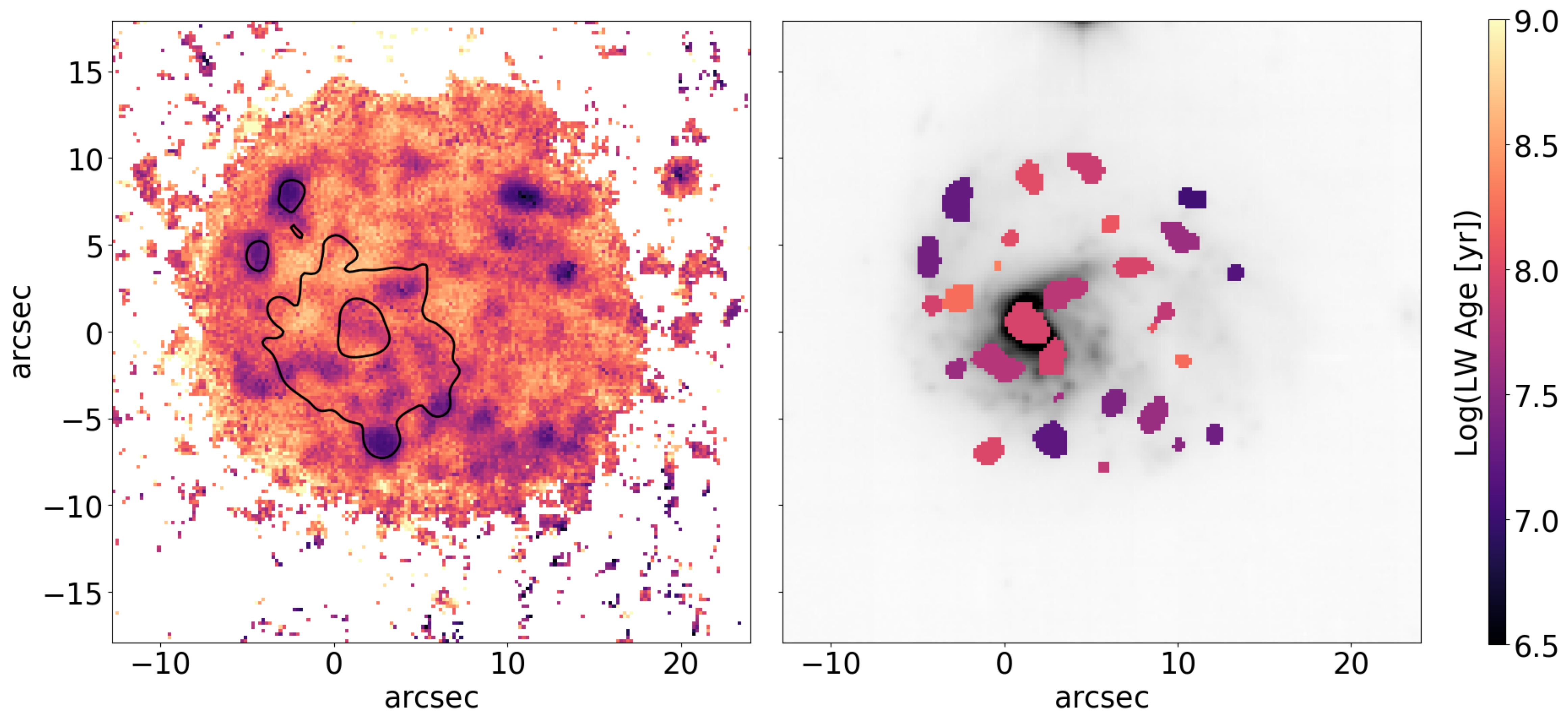}
\caption{Luminosity weighted age (LWA) map. {\it Left.} Contours represent the original body (see Fig.\ref{fig:SFH_maps}). {\it Right.} Only regions corresponding to the 29  \Ha knots identified in Fig.\ref{fig:Ha} are color coded according to the median LWA of each knots. In the background, the white-light image of the galaxy (Fig.\ref{fig:white}) is shown, for reference.  \label{fig:LWA}}
\end{figure*}

\begin{figure}
\centering
\includegraphics[scale=0.3]{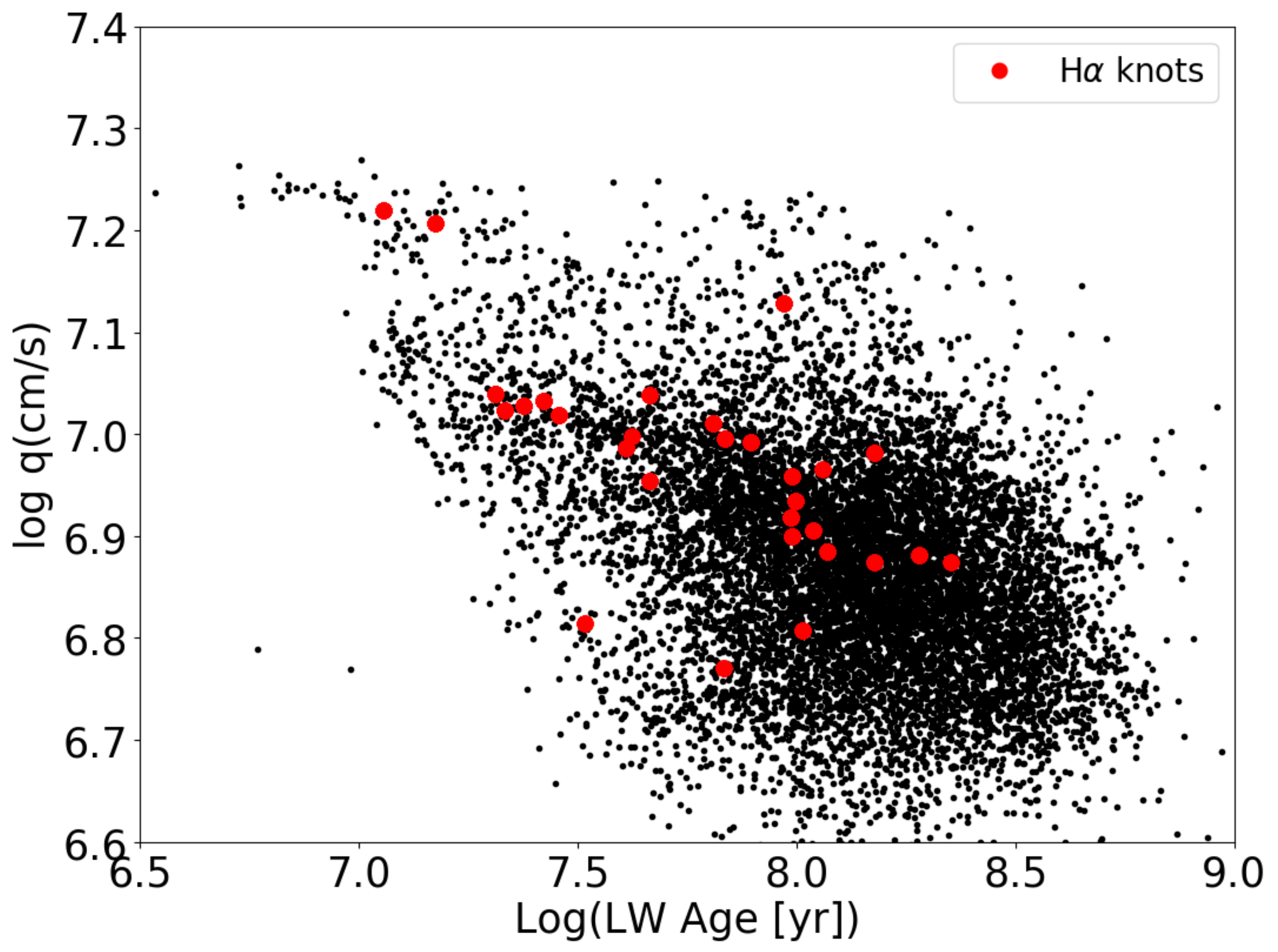}
\caption{Correlation between the luminosity weighted age  and the parameter of ionization. The average values computed in each \Ha knot are shown with red circles. \label{fig:LWA_q}}
\end{figure}

Figure \ref{fig:LWA} presents the map of the luminosity weighted age. This provides an estimate of the average age of the stars weighted by the light we actually observe, therefore giving us an indication on when the last episode of star formation occurred. The map  shows that the typical luminosity weighted age of the galaxy is $\sim 10^{8-8.5}$ yr. The distribution of ages is quite homogeneous,  except in some regions, where the stars are much younger ($\sim10^{6.5}$ yr). In most of the cases, these regions coincide with the \Ha knots, even though there are some knots with older ages and some young regions not included in any knot (see  the trail in the Southern part of the galaxy). Overall,  the knots have an age typically less than $10^{7.5}$ yr. The youngest knots are located in the South-West side of the galaxy, and two very young  ($\sim10^{6.5}$ yr) knots are located on the opposite side and others on the North-West side.

Comparing the luminosity weighted age and the ionization parameter  (Fig. \ref{fig:LWA_q}), we find a clear anticorrelation, both for the spaxels in the \Ha knots and for those outside. This  trend suggests that we might be able to infer the level of ionization in a galaxy given the age of its stellar population, or viceversa. Within the GASP sample, other galaxies present very different correlations. Understanding the physical meaning of this result is beyond the scope of this work and will be investigated in a following paper. 

\begin{figure*}
\centering
\includegraphics[scale=0.36]{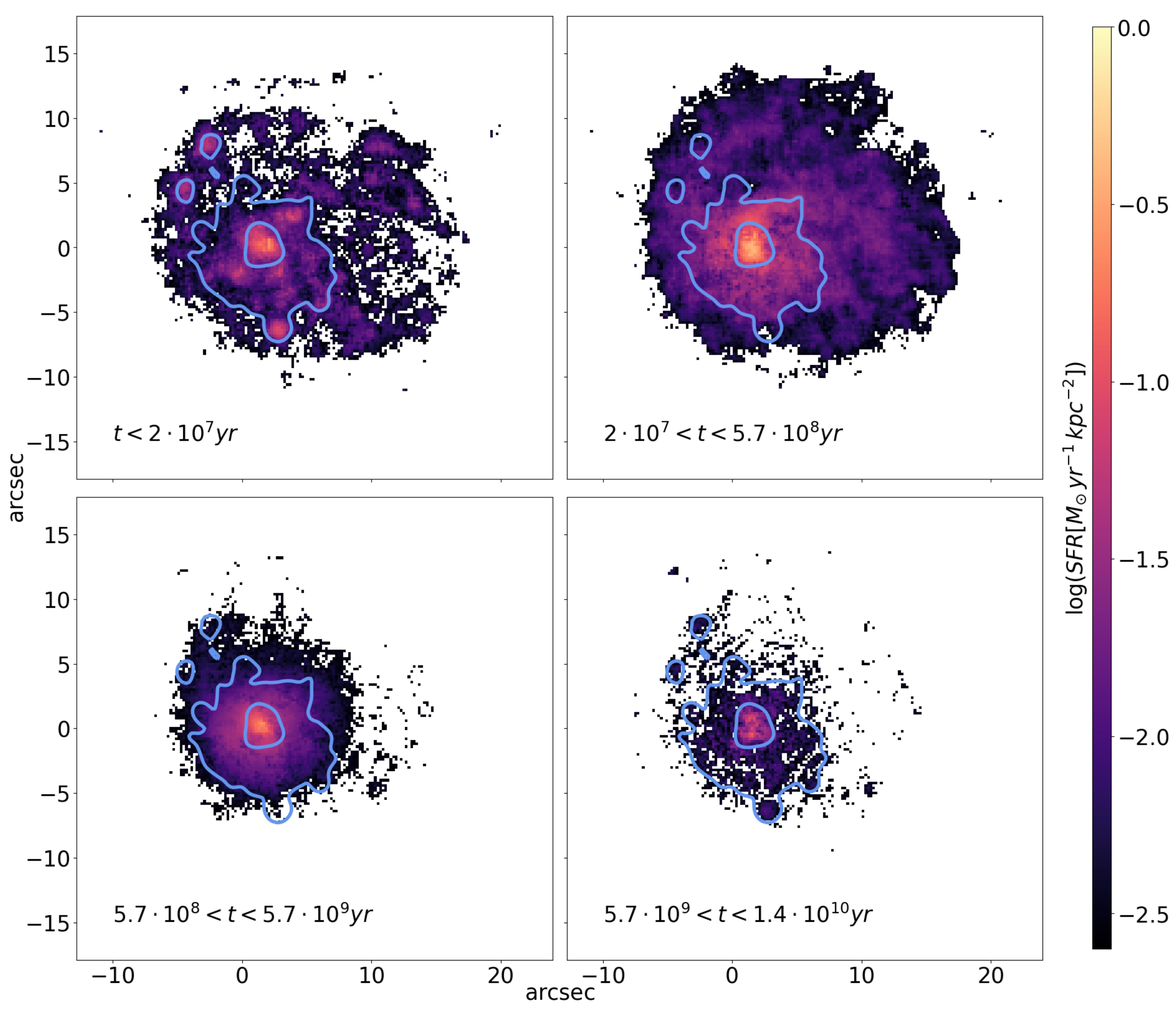}
\caption{Stellar maps of different ages, illustrating the average star formation rate per kpc$^2$ during the last $2\times 10^7$ yr (top left), between $2\times 10^7$yr and $5.7 \times 10^8$yr (top right), $5.7 \times 10^8$yr and $5.7 \times10^9$yr (bottom left) and $> 5.7 \times 10^9$yr ago (bottom right). Contours in all panels represent the original body, defined by the contours in the bottom right panel. \label{fig:SFH_maps}}
\end{figure*}

\begin{figure}
\centering
\includegraphics[scale=0.27]{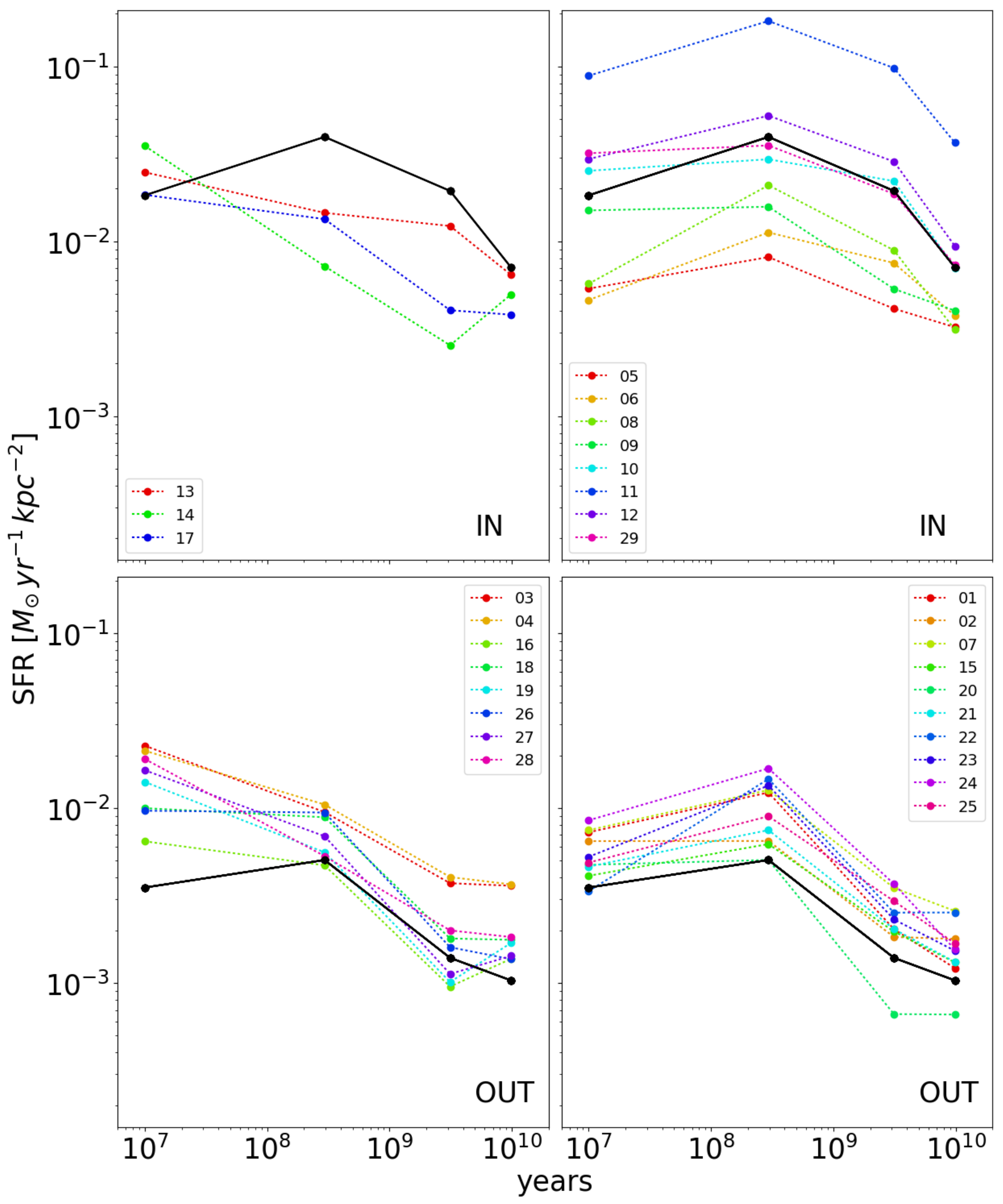}\caption{Star Formation histories (average SFR in four age bins) per unit kpc$^2$ within (upper) and outside (bottom) the galaxy original body (see Fig.\ref{fig:SFH_maps}), for each of the \Ha knots (colored dashed lines). For the sake of clearness,  the panels on the left show the \Ha knots with rising SFH, those on the right the \Ha knots with declining SFH. The SFH for the entire galaxy within and outside the original body (black solid lines) is also reported, for comparison.
\label{fig:SFH_tot}}
\end{figure}

Finally,  we can investigate  how the  SFR  varied with cosmic time.  We choose  four logarithmically spaced age bins in such a way that the differences between the spectral characteristics of the stellar populations are maximal (\citealt{Fritz2007} and \citetalias{Fritz2017}).

The bottom right panel of Fig. \ref{fig:SFH_maps} shows the SFR that took place in the oldest age bin ($t>5.7\times 10^9$ yr). This is the bin we use to define the contours indicating the original body shown in all the previous figures.  During this epoch, the SFR is high only within the original body, which assembled over this time. Very little SFR is found far from the galaxy core ($\sim 10^{\prime\prime}$). At this stage the galaxy does not show any sign of asymmetry. In the following age bin ($5.7\times 10^8$ yr$<t<5.7\times 10^9$ yr), the  galaxy SFR is still mainly concentrated in the original body and  the SFR in the outer region is still low. In the next  age bin ($2\times 10^7$ yr$<t<5.7\times 10^8$ yr), the SFR in the galaxy boosts and the galaxy develops especially towards the West side. This is the epoch when  the total SFR is the highest and the galaxy has the largest extension. The map is not smooth, suggesting the SFR is taking place in a non uniform way.  Finally, in the current SFR ($t<2\times 10^7$ yr), the core of the galaxy is still highly star forming, while in the external regions a decline in the SFR is observed. A conspicuous number of highly star forming regions are visible. Overall, they correspond to the knots seen in the \Ha maps.  

The Star Formation History (SFH), i.e. the average SFR in the four age bins, computed by summing the SFR in the spaxels, divided by the number of spaxels, is presented in Figure \ref{fig:SFH_tot} as black lines, for both the region within and outside the original body. The trends just discussed emerge clearly: the original body developed in the earlier epochs and the galaxy was forming stars at a quite high rate already in the second oldest age bin, while the external part developed in a later phase and a strong SFR increase happened only during the last $5.7\times 10^8$ yr.  Both regions  peak within $2\times 10^7$ yr$<t<5.7\times 10^8$ yr ago. The same Figure also shows  the SFHs of each \Ha knot separately. 
The SFHs of the \Ha knots within and outside the original body follow the trends of all the spaxels in the same regions, respectively, with some exceptions. Among the \Ha knots with rising SFR, 13, 14, and 17 (within the original body) and  16, 18, and 19  (outside) are all located in the South-West side of the galaxy. \Ha knots 3 and 4 are instead in the opposite side: in the North-East side. Finally,  \Ha knots 26, 27 and 28
are on the Western side, in the outskirts. The last four seem to be located on the extension of the same spiral arm.

This analysis shows how \Ha knots located at similar distances from the center but in different regions of the galaxy have a quite different history, suggesting they were born in different conditions. 

\section{Discussion}\label{sec:disc}
In this section we aim at drawing a scenario for the formation and evolution of \pp able to explain its main characteristics illustrated in the previous sections. First, we need to characterize the environment around the galaxy, in order to shed light on the possible physical mechanisms acting on it.

\subsection{The environment}

\begin{figure*}
\centering
\includegraphics[scale=0.4]{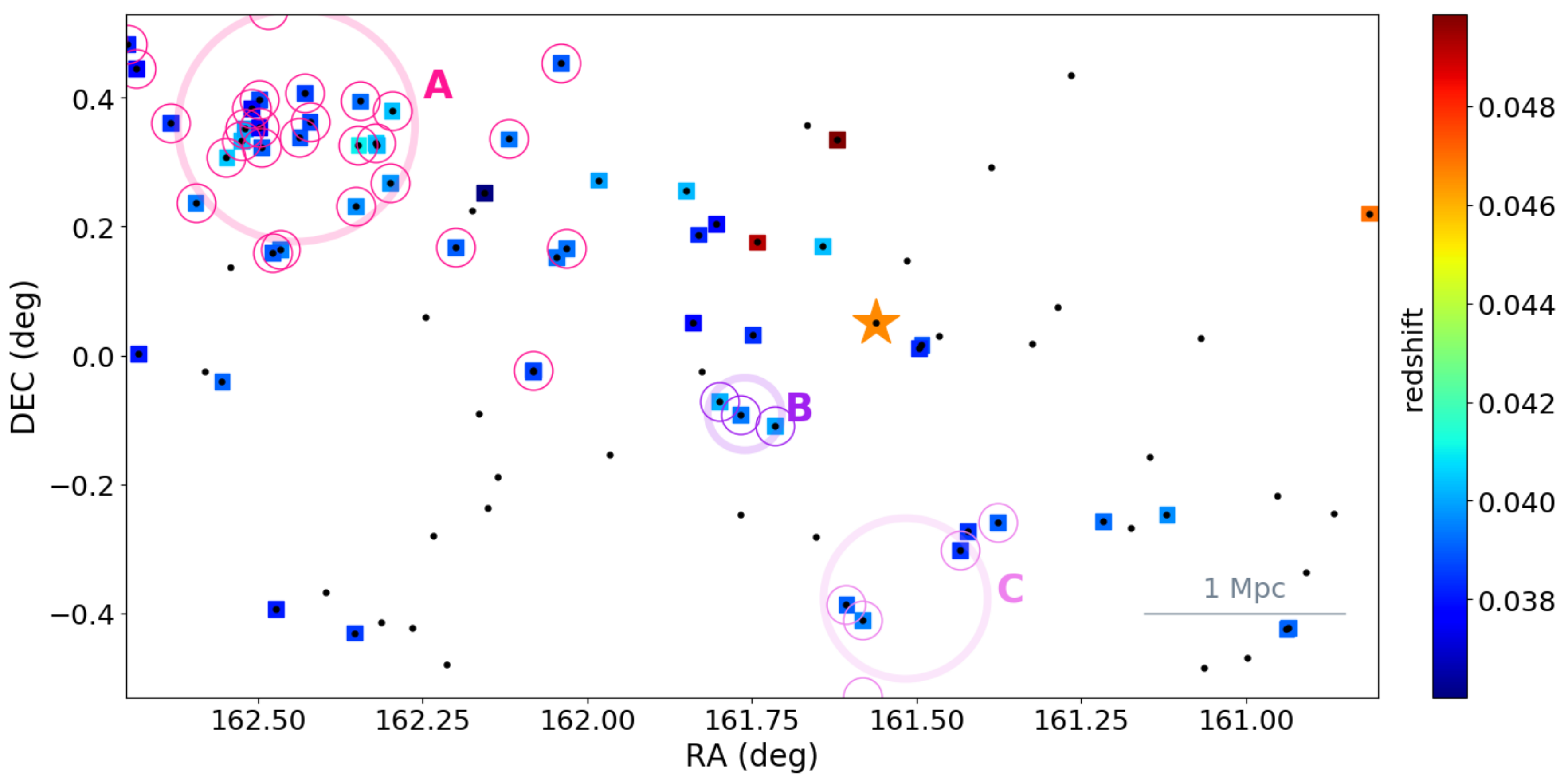}
\caption{ Position on the sky of galaxies around \pp in the redshift range $0.025<z<0.07$ (black dots). Galaxies in the redshift range $0.035<z<0.05$ are color-coded according to their redshift (squares). \pp is indicated with an orange  star. Galaxies belonging to groups identified by \citet{Calvi2011, Tempel2012} are highlighted with empty symbols (see Tab.\ref{tab:groups} for details). Shaded circles indicate the virial radius of the groups. The scale in the bottom right corner shows 1 Mpc at the redshift of \pp. \label{fig:env} }
\end{figure*}

\begin{table}
\caption{Properties of the groups around P11695 \label{tab:groups}}
\centering
\begin{tabular}{rrrrrrr}
\hline
  \multicolumn{1}{c}{ID} &
  \multicolumn{1}{c}{Ngal} &
  \multicolumn{1}{c}{z} &
  \multicolumn{1}{c}{RA} &
  \multicolumn{1}{c}{DEC} &
  \multicolumn{1}{c}{Rvir} &
  \multicolumn{1}{c}{$\sigma$} \\
  \multicolumn{1}{c}{} &
  \multicolumn{1}{c}{} &
  \multicolumn{1}{c}{} &
  \multicolumn{1}{c}{(J2000)} &
  \multicolumn{1}{c}{(J2000)} &
  \multicolumn{1}{c}{(Mpc)} &
  \multicolumn{1}{c}{($\rm km \, s^{-1}$)} \\
\hline
  A & 32 & 0.04036 & 162.44151 & 0.3568 & 0.51 & 245.8 \\
  B & 3 & 0.04106 & 161.76073 & -0.09061 & 0.16 & 118.1 \\
  C & 5 & 0.04008 & 161.51663 & -0.3767 & 0.35 & 134.1 \\
\hline\end{tabular}
 \tablecomments{Data taken from \citet{Tempel2012}. }
\end{table}

\pp does not belong to any recognizable cluster or group: it has been selected as isolated  from both our own environmental catalog \citep{Calvi2011} and  an independent catalog assembled by \cite{Argudo2015}.  Paccagnella et al. (in prep) estimate that  it is embedded in a halo of mass $M_\ast=3.9 \times 10^{11} \, M_{\odot}$.

Figure \ref{fig:env} shows the spatial distribution of all galaxies around P11695 with measured redshift in the range $0.025<z<0.07$. Redshifts are taken from the MGCz \citep{Driver2005} and SDSS-DR9 \citep{Ahn2012}. The estimated spectroscopic completeness  in this region of the sky is $>96\%$ at B=20 \citep{Liske2003,Driver2005}. 
In agreement to \cite{Argudo2015}, the closest galaxy at the same redshift is at 2.3 Mpc. \cite{Calvi2011, Tempel2012} identified overall three groups in this area, whose position is highlighted in Fig. \ref{fig:env} and whose properties are listed in Tab.\ref{tab:groups}. The spectroscopic galaxy sample adopted by \cite{Tempel2012} to identify the groups is complete  up to the Petrosian magnitude $m_r = 17.77$ \citep{Strauss2002}.

Their redshift difference with respect to \pp corresponds to a relative velocity of $\sim 2000$ $\rm km \, s^{-1}$.  This difference in velocity is $> \sim 10 \times$ the velocity dispersion of the largest group. Therefore, if \pp ever belonged to one of these  groups, it should now move at an unrealistic velocity. We therefore tend to exclude that the galaxy was formed in one of the groups and was subsequently kicked out from it. 

To firmly exclude the presence of massive structures around \pp, we searched for X-ray emission  within 30$^{\prime\prime}$ from the galaxy.  ROSAT (0.1- 2.4 keV) and SWIFT (15-25keV e 25-45keV, 50-194 keV) data are available  in this region of the sky, and both failed in detecting any extended structure, which could be connected to the presence of a group or cluster.

\pp is therefore a truly isolated galaxy and 
the typical mechanisms usually invoked to explain the observed tails, e.g. ram pressure stripping due to the interaction between the hot and dense intragroup/cluster medium and the galaxy gas \citep{Gunn1972}, or tidal interactions \citep{Byrd1990}, have to be ruled out, because they  generally take place in more massive or at least richer environments.  

In addition, no clear sign of recent merger is visible in any part of the analysis we have conducted, and there is no visible companion. Note that even if we do not observe companions now, they might have merged with P11695 in the past. \cite{Walker1996} show that perturbations induced by a minor merger can last for 1 Gyr. However, it is not clear whether a minor merger can induce significant lopsidedness for a sufficiently long time.  \cite{Walker1996} show that minor mergers can substantially heat and puff-up the disk, driving the formation of a strong bar ($m$ = 2 mode). Their simulated galaxies are not axi-symmetric at intermediate time during the merger, but they do not look lopsided after the satellites has been completely accreted. Similarly, \cite{Bournaud2005} find that a 10:1 minor merger triggers significant lopsidedness, but this lasts typically no more than  500 Myrs.

In contrast, fly-by events can trigger lopsidedness \citep[e.g.,][]{Mapelli2008}, but at much lower level than gas accretion. Moreover, in case of a fly-by we should still see the companion close to P11695.

Hence, we need to invoke some other mechanism to explain the peculiarities of \pp.

\subsection{Cold gas accretion scenario}
We propose a scenario in which
the galaxy is being fed by a filament of cold (T$\leq 10^5$ K) low metallicity gas, most probably flowing from the South West side of the galaxy. This accretion increases the availability of gas pumping up the lopsidedness of the disk, igniting the formation of giant HII regions, influencing the gas motion, the conditions of star formation and the metallicity of the galaxy. 

In our observations, there are many indirect parts of evidence that are consistent with and point to the cold gas accretion scenario, as we describe in what follows. 

\subsubsection{Lopsidedness}

\begin{figure}
\centering
\includegraphics[scale=0.43]{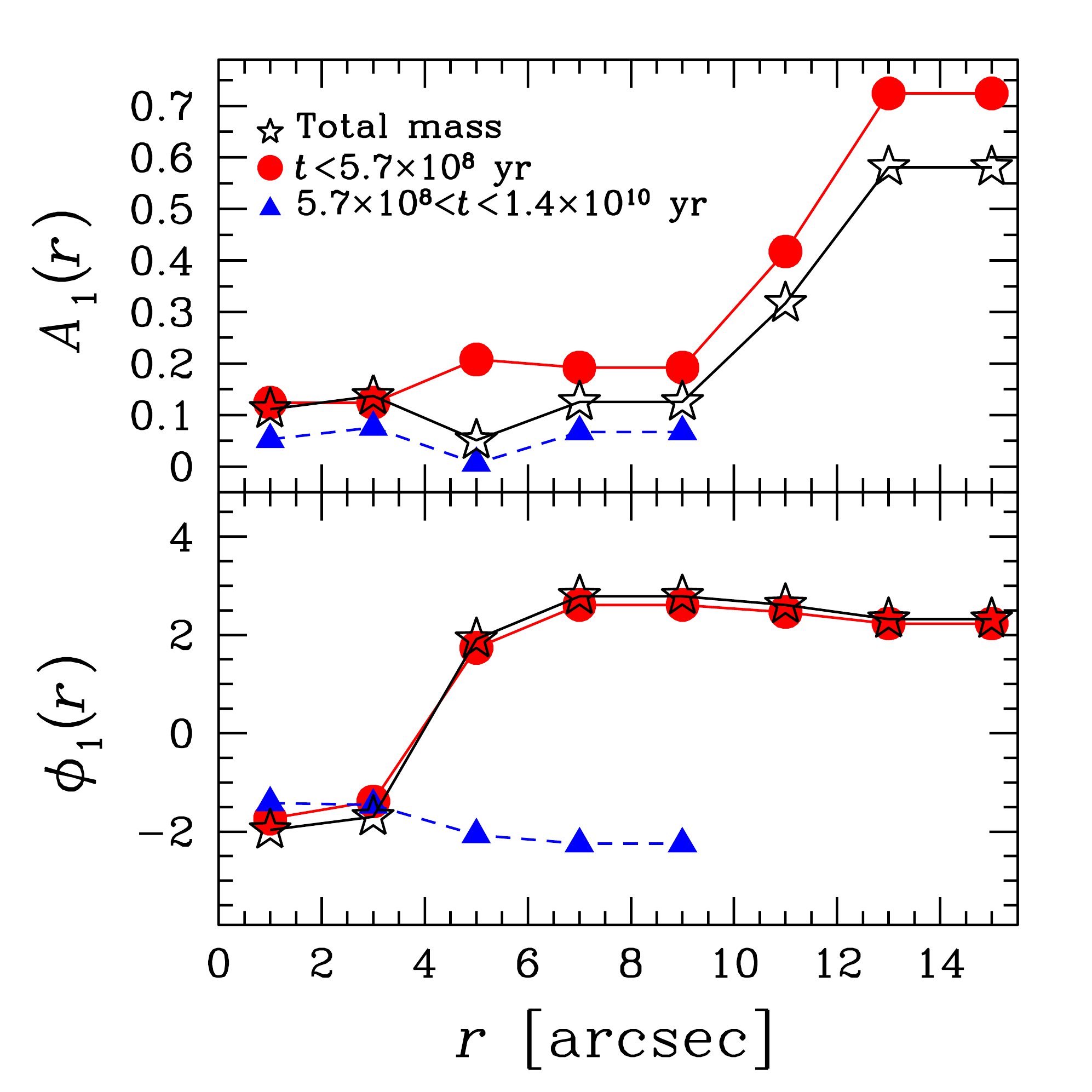}
\caption{Normalized amplitude of the disk lopsidedness ($A_1(r)$, top) and phase of the Fourier component $m=1$ ($\phi_1(r)$, bottom) as a function of the radius $r$, for the total mass (black stars), the mass formed $<5\times 10^7 yr$ ago (red circles) and the mass formed in the range $5.7\times 10^8 yr<t<1.4\times 10^{10} yr$  (blue triangles). \label{fig:A1} }
\end{figure}

\pp is a spiral galaxy with very asymmetric features. It is lopsided  and extends toward the West side much more than toward the East.

To quantify the lopsidedness in its surface density, we compute the Fourier  transforms of the stellar mass surface density, obtained from {\sc sinopsis}.

Following \cite{Bournaud2005}, the surface density is decomposed as:

\begin{equation}
\mu(r,\phi) = \mu_0(r) + \sum_m a_m(r) \cos (m \phi - \phi_m (r))
\end{equation}
where the normalized strength of the Fourier component $m$ is $A_m (r) = a_m /\mu_0 (r)$. Thus, $A_1$ represents the normalized amplitude of the disk lopsidedness at a given radius. Values of $A_1>0.1$ indicate significant lopsidedness. The quantity $\phi_m$ denotes the position angle or the phase of the Fourier component $m$. 

Figure \ref{fig:A1} shows the amplitude $A_1\,{}(r)$ and the phase $\phi_1\,{}(r)$ versus radius $r$ for the total stellar mass and the mass formed in two age bins ($t<5.7\times 10^8$ yr and $5.7\times 10^8 $yr$<t<1.4\times 10^{10}$ yr). Considering the total mass, \pp appears to be significantly lopsided for $r>10^{\prime\prime}$. The phase $\phi_1$ is nearly constant for $r>5^{\prime\prime}$, which is typical of most lopsided galaxies \citep[see, e.g.,][]{Rix1995}.  Contrasting the curves for the two age bins,  we find an increase in both the amplitude and the phase of  the lopsidedness with time.  This suggests that the West side of the galaxies developed in recent epoch, $t<5\times 10^7$ yr ago, in agreement with what found in Fig.\ref{fig:SFH_maps}.

In principle, the asymmetric shape could be due to either  accretion or a disk truncation, after a merger or due to a gas stripping.  However, if the lopsidedness was due to mergers, we should also detect signs of the the remnants in our spatially resolved spectroscopy \citep{Mapelli2008}, which is not the case. As for ram pressure, the galaxy should be located in a denser medium to feel the gas stripping, while we find no observational support for this hypothesis. In contrast, no observational evidence contrasts the gas inflow.

The simulations run by  \cite{Bournaud2005}  show that lopsidedness can result from cosmological accretion of gas on galactic disks, which can create strongly lopsided disks when this accretion is sufficiently asymmetrical. Simulations by e.g. \cite{Brook2008, Roskar2010, Snaith2012} predict that misaligned gas structures are found to form and persist for many Gyr through the continued accretion of cold gas with misaligned angular momentum. In absence of any continuous gas fuel, in an axisymmetric potential the gas will over time relax back into one of the preferred axes, becoming exactly co- or counter-rotating with the stars \citep[see also][]{vandeVoort2015}.

This analysis therefore supports a scenario in which continuous gas accretion is feeding the galaxy inducing the growth of the West side of the galaxy.  Based on the stellar population analysis, we can date the beginning of the galaxy growth due to gas accretion  $<5.7\times 10^8$ yr ago.

\subsubsection{Velocity profiles}
\cite{Jog1997, Jog2002} showed that a galaxy showing spatial asymmetry would naturally show kinematical asymmetry. The two have to be casually connected in most cases. Instead, the gas and stellar components co-rotate and span the same velocity range in \pp. The gas kinematic shows a bending of the locus where velocities are negative  in the external regions of the galaxy. It convexity points toward East. 
Nbody/hydrodynamical simulations have shown that this U-shape might be due to ram pressure stripping \citep{Merluzzi2016}, but it might also be related to galaxy interactions or gas accretion that produce lopsidedness \citep[see,  e.g., ][and references therein]{Sancisi2008}. In \pp, the bending is consistent with a scenario in which the gas is inflowing from South-West with its own velocity, and affects the gas velocity in its motion toward the center. The stellar kinematics might show a similar pattern, but this is less evident because this component is more bound than the gas, and therefore less easily disturbed.

The maps of velocity dispersions show that gas and stars have different level of turbulence. Overall, the velocity dispersion of the gas decreases from the center towards the outskirts. 
In contrast, the stars have a more quite motion in the core of the galaxy, but the velocity dispersion significantly increases toward the external regions. The SFH maps (Fig. \ref{fig:SFH_maps}) have shown that the galaxy outskirts have formed in a recent epoch and the motion of the newly formed stars might be influenced by the relative velocity and inclination of the inflowing gas.

\subsubsection{Metallicity gradients}

\pp has a generally low metallicity 
and  a strong metallicity gradient. Compared to other gradients of late-type galaxies in the local universe \citep{Pilyugin2014,Ho2015}, \pp has one of the steepest ones. Overall, the metallicity is higher in the core, in the North-West and South-East sides of the galaxy, while it is significantly lower in the South-West and North-East sides.  

\cite{Mapelli2008, Oppenheimer2010, Ma2016} suggest that strong metallicity gradients might support the cold gas accretion scenario: when particularly low metallicity gas is detected in a galaxy halo it is often claimed to be the accretion of the IGM filaments, because the pristine gas accreted through filaments is expected to be characterized by a lower metallicity than that of the target galaxy \citep[see also][for higher $z$ results]{Crighton2013, Lehner2013, Cooper2015}.

\cite{Koppen2005} investigate the chemical evolution of galaxies that undergo an episode of massive and rapid accretion of metal-poor gas with models using both simplified and detailed nucleosynthesis recipes. Focusing on the effect of gas accretion on metallicity gradients in the last 2 Gyr, they find that during the infall, the oxygen abundance is decreased due to dilution of the galactic gas, followed after the infall by the evolution towards the closed-box relation. The large excursions in the gradient is reproduced if the mass of the infall is much larger than the mass of the gas present in the galaxy and the infall is greater than the SFR.

Radial metallicity gradients are instead expected to flatten once previously-ejected gas begins to re-accrete. This is because metals that form and are ejected in a galaxy's core are mixed by the halo and re-distributed to large radii. 
We should therefore be able to exclude a recycle of gas in \pp.

In principle, we could expect the region where the gas is inflowing to be much more metal poor than the rest of the galaxy, with the result of having much more asymmetric gradients on the different sides of a same slit. This is true if the low metallicity gas is able to form stars before having traveled across the galaxy and mixing up with the existing gas. 
We can therefore estimate the time a particle would need to complete an orbit around the galaxy. As a first approximation, the orbital period is $P=2\pi\frac{r}{v_c}$ with $r$ radius of the galaxy, and $v_c$ rotation velocity. From Fig.\ref{fig:prof_vel} we obtain $v_c\sim 50 km/s$ for $r>3^{\prime\prime}\sim=$2.7 kpc. We adopt as inclination $i=42$ deg. The orbital period results to be $\rm{\sim 2.5\times 10^8 \, yr}$ for $r=3$ kpc,  $\rm{\sim 8\times 10^8 \, yr}$ for $r= 10$ kpc. As in the North-East side of the galaxy we do not observe strong metallicity gradients, the orbital periods suggest that the gas did not have the time yet to complete an entire orbit and influence the metallicity of all new born stars, and therefore accreted in a later epoch. 
 We only measure relatively low metallicity in the South-West and North-East sides of the galaxy. The former is the region where we hypothesize the gas is inflowing, i.e along the probable major axis of the original body of the galaxy. 

\begin{figure*}
\centering
\includegraphics[scale=0.43]{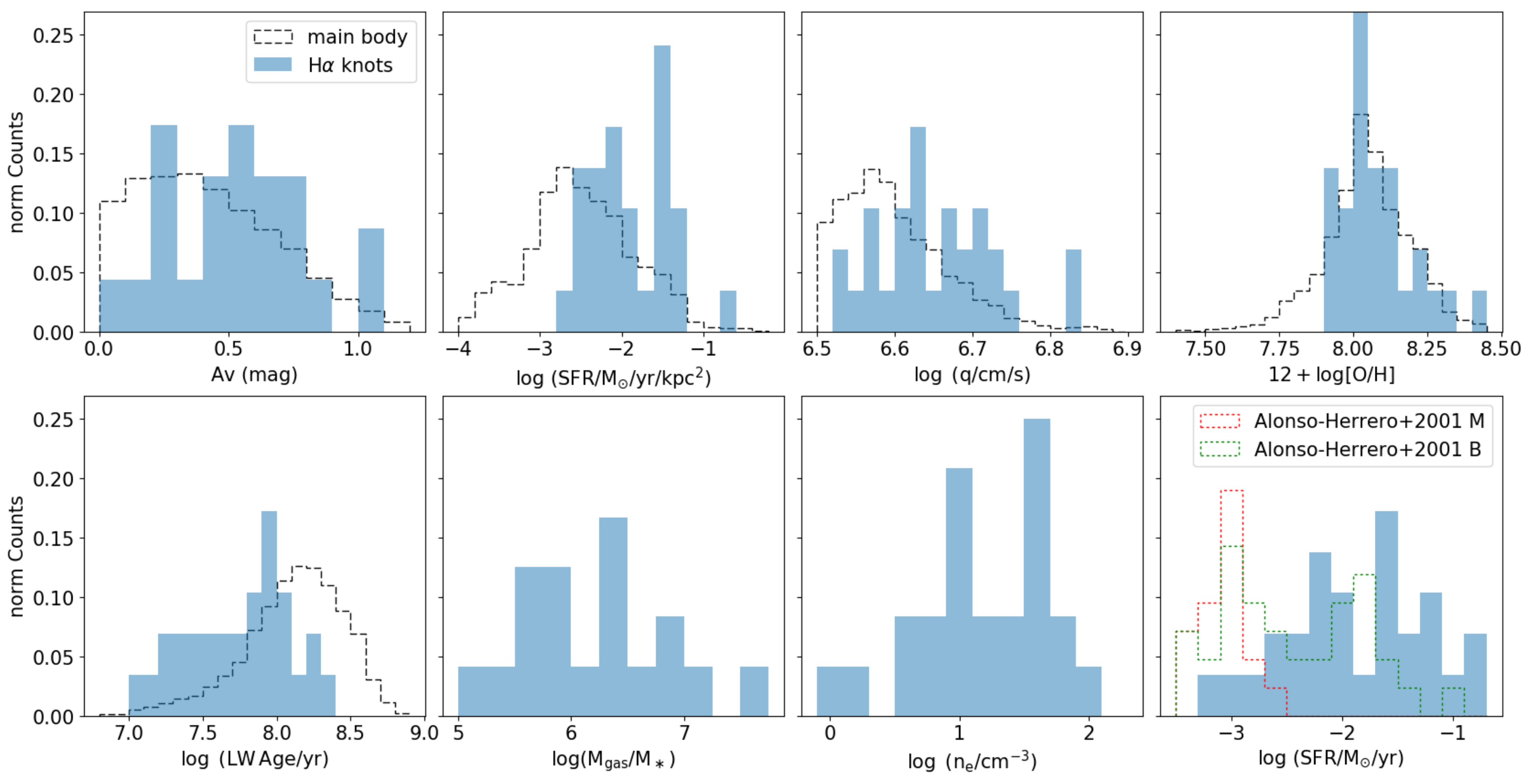}
\caption{Normalized distribution of the properties of the \Ha knots, compared to that of those of the galaxy main body (when meaningful). The $A_V$, SFR per unit kpc$^2$, parameter of ionization, metallicity, luminosity weighted age, gas mass,  electron density and SFR are shown. In the bottom right panel, we report for comparison the SFR distributions obtained from the \Ha luminosity as measured by  \citet{Alonso2001} for HII regions in nearby galaxies.  Estimates using only the brightest (B) and median (M) \Ha luminosities are shown. \label{fig:blob_distr} }
\end{figure*}

\subsubsection{\Ha knots}
Another peculiarity of \pp is the presence of many \Ha knots. They are all on the main body and present elongated shapes.

Figure \ref{fig:blob_distr} shows the distribution of the properties in the knots compared to those in the main body. Median values obtained by considering all the spaxels within each \Ha knot agree with those obtained from the spectra integrated over the knots.  
Overall, the knots are dustier, have higher SFR values,  younger ages, higher ionization parameters than the rest of the galaxy, and similar  metallicity. A gradient in the properties of the knots, going from North East to South West, might be detected in all the properties, with the exception of the 2 more external knots in the North West area. Half of the current SFR is taking place in the \Ha knots, which are therefore an important hotbed of new stars.
In the last panel of Fig. \ref{fig:blob_distr}, we overplot the SFR distribution of HII regions of a sample of 52 nearby galaxies, studied by \cite{Alonso2001}. These authors provide the brightest, the median and the faintest \Ha luminosities of the circumnuclear HII regions.  We converted their luminosities to SFRs and find that indeed the \Ha knots in \pp are systematically larger than both the typical and brightest HII regions in other galaxies.

Considering the SFH of the \Ha knots, we find that most of them follow the trend of the entire galaxy, except for the \Ha knots located in the South-West region. They have all rising SFHs, suggesting that a large amount of fuel for star formation is available in this region of \pp.

Even though clumpy galaxies in the local universe  are not as common as at higher redshift \citep[e.g.,][]{Genzel2006, Agertz2009, Elmegreen2007}, samples at  redshifts similar to that of \pp have been studied e.g. by \cite{Smith2001, Garland2015}. Overall, the continuous appearance of clump-cluster galaxies throughout a wide range of redshifts means that disk galaxies start forming over an extended period of time.

We are not able to state whether the knots we  observe are the first knots formed in the galaxy or if their are a n$^{th}$ generation of knots. For example, \cite{Elmegreen2007} claim that the clump- cluster phase lasts for  0.5-1 Gyr and occurs only once during the life of a galaxy, unless there is a significant accretion event later. In contrast, simulations by \cite{Noguchi1999} do not exclude the possibility that new generations of knots are easily formed.

The explanation for the existence of these knots is that when a disk becomes sufficiently massive through accretion (through cold gas inflow in our case), giant clumps form at the local Jeans mass by gravitational instabilities. This induces a disk fragmentation,  without destroying the disk. The mass of the clumps therefore depends on the turbulent Jeans mass. As turbulent speeds decrease relative to the rotation speed, clump masses decrease relative to the galaxy mass; then their interactions and migrations toward the bulge become less severe.

As described in detail in \citetalias{Poggianti2017a}, we can compute the gas mass of each star forming region from the \Ha luminosity \citep{Boselli2016}, after having estimated the electron density. To do that, we follow the relation presented in \cite{Proxauf2014}  that is based on the R =[SII]6717/[SII]6737 line ratio and is valid in the range 0.4 < R < 1.435.  
24 out of 29  knots have a [SII]6716/[SII]6732 ratio in the range where the density calibration applies, while the others have ratio values larger than 1.44, which suggests that their density is below 10 cm$^{-3}$. The distribution of their ionized gas densities is shown in Fig. \ref{fig:blob_distr}. Most of the measured densities are between 3 and 100 cm$^{-3}$, with a median of 19 cm$^{-3}$. The derived  ionized gas mass distribution is also shown in Fig. \ref{fig:blob_distr}. Most of the knots have masses in the range $10^5-10^{7.5} M_\odot$, with a median of 7.9$\times10^5 M_\odot$. Summing up the gas mass in these knots we obtain $\sim 10^8 M_\odot$. This value has to be taken as lower limit to the total ionized gas mass, given that the contributions of the knots with no density estimate and of the diffuse line emission are not taken into account.

Note that if major interactions were occurring, the resulting galaxy would likely include a highly concentrated central starburst, not smaller clumps of star formation spread throughout.

\subsubsection{Star forming properties} 
The typical luminosity weighted age of \pp is $10^{8-8.5}$ yr, but the galaxy is characterized by a large number of much younger regions ($\sim 10^7$ yr). In particular, Figure \ref{fig:LWA} suggests the existence of a trail of very young stars starting from the South-West side, extending toward East and  continuing on the North side toward West. It is tempting to interpret this as the path of new gas inflowing and then orbiting around the original body of the galaxy. 

The analysis of the SFHs showed that  at the early stage of its formation,  \pp probably was a much more symmetric object. The lopsidedness seems to have started to develop at later times, especially in the second youngest bin.
The measurement of the amplitude of the lopsidedness confirms this finding. In the  youngest age bin the SFR is declining, and the outskirts of the galaxies form stars at a lower level.
Also these findings are consistent with a scenario according to which after its formation, \pp has been fed by a gas inflow, leading to a more massive extension of the galaxy on the West side.  The decrease in star formation observed today might suggest that currently the gas flow is feeding the galaxy at a lower rate.

\subsection{Final remarks}
\pp has been observed by the ALFALFA survey \citep{Giovanelli2005}.\footnote{The galaxy is not in the publicly released ALFALFA catalog, since its S/N is below the adopted limit (S/N>6).} Given the  low resolution of the observations, we can not characterize the lopsidedness, but we can obtain an estimates of its HI mass, which can be considered as the raw fuel for star formation and a lower limit of the accreted gas mass, as a large fraction of it has already formed stars. The inferred $M_{\rm HI}$ is $1.03\pm 0.2  \times 10^{10} M_\odot$ (D. Stark, priv. comm.), entailing a  $M_{\rm HI}/M_\ast \sim 0.8$. Comparing its position on the  $M_{\rm HI}/M_\ast$ vs. $M_\ast$ plane to that of galaxies at $z=0$ drawn from the RESOLVE survey \citep{Stark2016}, we find it lays on the upper envelope of the relation \citep[see also][]{Putman2017}, suggesting its ratio is higher than the typical one for galaxies with similar stellar mass.   \pp has therefore still a large reservoir of unprocessed gas.   

Considering the SFR of the galaxy (\rm 3.27 $M_\odot \, yr^{-1}$), the star formation efficiency ($\rm SFR/M_{gas}$) turns out to be 0.31 $\rm Gyr^{-1}$, which gives us a time scale of the gas consumption. \cite{Bournaud2005} simulate accretion rates of the order to $\rm \sim 3-9 M_\odot \, yr^{-1}$ for $\sim$2 Gyr \citep[see also][]{vandeVoort2012}, yielding a total accretes mass of $\rm 0.6-1.8\times 10^{10} \, M_\odot$, i.e. the same order of magnitude as our estimates.
For comparisons, in the Milky Way,  gas that is clearly infalling is observed at 1-15 kpc above the disk. The actual rate of accretion depends on the 3D motions of the gas and the full extent of the accreting layer, but the rates calculated are $\rm \sim 0.1-0.4 \,  M_\odot \, yr^{-1}$ for the coldest gas \citep{Putman2012}, and closer to $\rm \sim 1 \,M_\odot \, yr^{-1}$ when the ionized gas is included \citep{Lehner2011}. In M33, a  direct detection of gas accretion yields an accretion rate obtained $\rm \sim 2.9 \, M_\odot \, yr^{-1}$. This amount is relatively large for this small galaxy, and may be further evidence for the infall of fuel being intermittent in nature \citep{Zheng2017}.

Taking into account the adopted inclination of \pp,  we can suppose an U-shaped warp for the galaxy visible in the most external regions. The possible presence of the warp could be indicated also by the fact that the velocity field indicates a change of the apparent major axis from the inner to the outer region of the galaxy. This velocity distribution can be represented by annular rings in circular motion, progressively more inclined at larger radii \citep{vanderKruit1978}.
\cite{Sancisi1976, Binney1992} have emphasized that warps occur in apparently isolated galaxies. They conclude that if the generator of the warps is invisible, this could be outer gas accretion \citep[see also][for more recent results]{Lopez2002, Bournaud2005,Kamphuis2013}, consistent with our general interpretation.

\cite{Reichard2009} connected the lopsidedness of 25,000 star-forming galaxies from SDSS to metallicity. They found that at a fixed mass, the more metal-poor galaxies turn out to be more lopsided, extending the morphology-metallicity relation to the full population of star-forming galaxies. This result might explain why \pp lay off the \cite{Tremonti2004}'s mass metallicity relation. \cite{Reichard2009} interpret their result in the context of a gas accretion triggering scenario. Indeed, the accreted gas is metal-poor \citep[e.g.,][]{Dekel2009, vandeVoort2012}, and  induces off-center giant star-forming clumps that gradually migrate toward the disk centers \citep{Ceverino2010, Mandelker2014}. The giant star-forming clumps may be born in-situ (if the accreted gas builds up the gas reservoir in the disk to a point where disk instabilities set in and trigger star formation) or ex-situ (if already formed clumps are incorporated into the disk).  In both cases, a significant part of the star formation in the disks occurs in these giant clumps. As a result of the whole process, the gas accretion produces bright off-center starbursts increasing the lopsidedness of the host disk.  This explanation suits also  our observational results.

Unfortunately, we are not able to get direct observations of this gas flow. 
As  discussed e.g. by \cite{Rubin2016},  an unequivocal evidence of gas flow toward a  source is the detection of absorption lines in a galaxy's spectrum which shows a velocity shift with respect to its rest frame. The continuum spectrum can be either a higher redshift background source or the host galaxy itself.  This technique is sensitive to the inflow of material over a broad range of densities and temperatures. 
Even though we found a high redshift ($z= 0.458$) galaxy in the MUSE FoV, $\sim 40^{\prime}$ NW from \pp, its flux is too weak to detect any possible sign of absorption by the gas inflow. 

\section{Summary and conclusions}

GASP (GAs Stripping phenomena in galaxies with MUSE) is an ongoing ESO Large Program with the MUSE/VLT to study the causes and the effects of gas removal processes in galaxies in different environments. Within the sample, we identified a galaxy that is likely undergoing accretion
 through a cold gas filament coming from the South-West side of the galaxy. 
In this paper we analyze its spatially resolved properties that indirectly support this scenario. 

\pp is an isolated spiral galaxy at $z=0.04648$ showing a marked lopsidedness in the light distribution. Such lopsidedness developed $<6\times 10^8$ yr ago.  The gas and the stars are co-rotating around the same axis, but the velocity field of the gas bends in the outskirts, as if new gas has been infalling with a different orientation and velocity.

\pp  presents steep metallicity gradients, as expected from simulations when cold accretion of low metallicity gas is occurring \citep[e.g.,][]{Mapelli2008}. 
\pp has low levels of dust ($A_V<1$mag), a low ionization parameter value ($\log(q)\sim$ { 6.8}) and quite young  luminosity weighted age ($\rm{\log(LWA)\sim 8.4}$).  Both the luminosity weighted age and parameter of ionization maps are very patchy, and we found a close correspondence between the two. A trail of very young stars ($\rm{\log (LWA)\sim7}$) visible across the galaxy might trace the path of the new gas.
Another important characteristic is the presence of 29 elongated \Ha knots on the galaxy disk. These knots have systematically lower metallicity, younger ages, higher ionization parameters than the rest of the galaxy. 

Finally, it is worth stressing again that this galaxy was selected for showing signs that could be indicative of stripping, therefore it emerges that in optical images gas stripping and gas accretion can present similar features. 
Within the GASP sample, we recently detected other
galaxies not belonging to any structure whose properties might resemble those of \pp. In a following paper we will characterize this whole sample, in order to state whether all isolated galaxies that 
present signatures suggestive of gas stripping
are on the contrary still accreting gas.

\acknowledgments
We thank the anonymous referee whose comments helped us strengthening the manuscript.
Based on observations collected at the European Organisation for Astronomical Research in the Southern Hemisphere under ESO programme 196.B-0578. 
This work made use of the {\sc kubeviz} software which is publicly available at \url{http://www.mpe.mpg.de/~dwilman/kubeviz/}. 
We thank Marco Ajello, Valentina La Parola and Fabrizio Nicastro for useful discussions. We thank the ALFALFA team for sharing their private catalog and  David Stark for providing us with the HI measurements and for stimulating discussions.  
We are grateful to Joe Liske, Simon Driver and the whole MGC collaboration for making their dataset easily available, and to Rosa Calvi for her valuable work on the PM2GC. 
We acknowledge financial support from PRIN-INAF 2014. B.V. acknowledges the support from 
an  Australian Research Council Discovery Early Career Researcher Award (PD0028506).  MM acknowledges financial support from the Italian Ministry of Education, University and Research (MIUR) through grant FIRB 2012 RBFR12PM1F, from INAF through grant PRIN-2014-14 (Star formation and evolution in galactic nuclei), and from the MERAC Foundation. J.F. acknowledges financial support from UNAM-DGAPA-PAPIIT IA104015 grant, M\'exico.
This work was co-funded under the Marie Curie Actions of the European Commission (FP7-COFUND). 

\facilities{VLT(MUSE)} 
\software{KUBEVIZ, ESOREX, SINOPSIS, IRAF, CLOUDY, pyqz, IDL, Python}

\bibliographystyle{yahapj}
\bibliography{gasp}


\end{document}